\documentclass[traditabstract]{aa}

\usepackage{graphicx}
\usepackage{txfonts}
\usepackage{rotating}
\usepackage{natbib}

\def\d{{\rm d}}
\def\pd{\partial}

\begin{document}

\title{Dynamics of multiple protoplanets embedded in gas/pebble disks
and its dependence on $\Sigma$ and $\nu$ parameters}
\titlerunning{Dynamics of multiple protoplanets embedded in gas/pebble disks}
\authorrunning{M.~Bro\v z et al.}

\subtitle{}

\author{M.~Bro\v z\inst{1}, O.~Chrenko\inst{1}, D.~Nesvorn\'y\inst{2}, M.~Lambrechts\inst{3}}

\institute{Institute of Astronomy, Charles University, Prague, V Hole\v sovi\v ck\'ach 2, 18000 Prague 8, Czech Republic,\\
e-mail:
mira@sirrah.troja.mff.cuni.cz
\and
Department of Space Studies, Southwest Research Institute, 1050 Walnut St., Suite 300, Boulder, CO 80302, USA
\and
Lund Observatory, Department of Astronomy and Theoretical Physics, Lund University, Box 43, 22100 Lund, Sweden
}

\date{Received ???; accepted ???}

 
\abstract{%
Protoplanets of Super-Earth sizes may get trapped in convergence zones
for planetary migration and form gas giants there.
These growing planets undergo accretion heating, which triggers a hot-trail
effect that can reverse migration directions, increase planetary eccentricities
and prevent resonant captures of migrating planets (Chrenko et al. 2017).
In this work, we study populations of embryos accreting pebbles, under different
conditions, by changing the surface density, viscosity, pebble flux, mass, and
the number of protoplanets. For modelling we use \textsc{Fargo-Thorin} 2D hydrocode
which incorporates a pebble disk as a 2nd pressure-less fluid,
the coupling between the gas and pebbles
and the flux-limited diffusion approximation for radiative transfer.

We find that massive embryos embedded in a disk
with high surface density ($\Sigma = 990\,{\rm g}\,{\rm cm}^{-2}$ at $5.2\,{\rm au}$)
undergo numerous `unsuccessful' two-body encounters which do not lead to a merger.
Only when a 3rd protoplanet arrives to the convergence zone,
three-body encounters lead to mergers.
For a low-viscosity disk ($\nu = 5\times10^{13}\,{\rm cm}^2\,{\rm s}^{-1}$)
a massive coorbital is a possible outcome,
for which a pebble isolation develops and the coorbital is further stabilised.
For more massive protoplanets ($5\,M_\oplus$), the convergence radius is located
further out, in the ice-giant zone. After a series of encounters, there is an evolution
driven by a dynamical torque of a tadpole region, which is systematically
repeated several times, until the coorbital configuration is disrupted and planets merge.
This may be a pathway how to solve the problem that coorbitals often
form in simulations but they are not observed in nature.

In contrast, the joint evolution of 120 low-mass protoplanets ($0.1\,M_\oplus$) reveals
completely different dynamics. The evolution is no longer smooth, but rather a random walk.
This is because the spiral arms, developed in the gas disk due to Lindblad resonances,
overlap with each other and affect not only a single protoplanet but several in the
surroundings. Our hydrodynamical simulations may have important implications for N-body
simulations of planetary migration that use simplified torque prescriptions and
are thus unable to capture protoplanet dynamics in its full glory.
}

\keywords{hydrodynamics -- protoplanetary disks -- planet-disk interactions -- planets and satellites: formation}

\maketitle


\section{Introduction}

A giant planet core formation is a key to understand the evolution
of the Solar System, and exoplanetary systems as well.
It is believed that gas giants, and possibly (some subset of) the super-Earths,
form in a special region of the protoplanetary disk
where an opacity transition (snowline) is located,
and a transition between the viscously-heated and irradiated (flared) parts
creates a suitable convergence zone \citep{Bitsch_etal_2014A&A...570A..75B}.
A relatively massive solid core of about $10$ to $20\,M_\oplus$
is then needed to accrete the surrounding gas \citep{Pollack_etal_1996Icar..124...62P},
before the disk dispersal.
Consequently, there is an ongoing `quest' for the fastest mechanism
which would beat other (slower) mechanisms of core formation.
At the same time, it is necessary to address the gas flow
which determines the actual critical mass.
A number of processes were already discovered
which contribute either in a positive way, e.g.
the pebble accretion due to an aerodynamic drag in the respective Hill spheres
\citep{Lambrechts_Johansen_2012A&A...544A..32L},
or a negative way, e.g.
the concurrent inflow and outflow in advective atmospheres
\citep{Lambrechts_Lega_2017A&A...606A.146L}.

In our previous paper \citep{Chrenko_etal_2017A&A...606A.114C},
we studied yet another process called the hot-trail effect
which arises naturally due to the heat liberated by the accretion of pebbles
onto protoplanets with several Earth masses.
This heat expands the gas behind the protoplanet,
creating an underdense region and changing the gravitational torque
of gas acting on the protoplanet.
In our setup, we simply assumed that the protoplanets radiate
the accretion energy and heat the surrounding gas;
there is no blanketing by the atmosphere,
or a magma ocean which would keep the heat within the protoplanets.
We used a self-consistent 2D model with interacting gas and pebble disks
(see below), so the pebble flux onto a given protoplanet is not simply prescribed.
As a result, the migration rates of protoplanets (or equivalently the torques)
are altered \citep{Benitez_etal_2015Natur.520...63B},
and their eccentricities increase substantially.
These non-zero eccentricities 'change the game',
because the captures in mean-motion resonances between protoplanets
are then much less probable.

\cite{Eklund_Masset_2017MNRAS.469..206E} used a 3D model to study
a dependence of the hot trail on the initial values of the eccentricity~$e_0$,
inclination~$i_0$, the protoplanet mass~$M_{\rm em}$, and the accretion rate~$\dot M$,
which was prescribed in their model. It turns out that asymptotic values
of the eccentricity~$e_{\rm asy}$ (or $i_{\rm asy}$) are important for the dynamics,
probably more than $\d e/\d t$, $\d i/\d t$. They also realized the inclination
can remain low ($<10^{-4}$\,rad) -- if its initial value was very low --
because even a moderate eccentricity suppresses a further increase of~$i$.
As a consequence, 2D models in which no vertical hot trail is present
may be still a viable and less expensive alternative.
Nevertheless, we shall keep track of orbital inclinations
excited by mutual encounters, because they do affect the rate
of pebble accretion when the protoplanets orbit above or below the pebble disk
\citep{Levison_etal_2015Natur.524..322L}.

Given the expensiveness of hydrodynamical computations, we considered
a single set of parameters in~\cite{Chrenko_etal_2017A&A...606A.114C},
although the dependence on disk(s) parameters is crucial.
In particular, different values of
the surface density~$\Sigma$,
the pebble flux~$\dot M_{\rm p}$, or
the viscosity~$\nu$
can potentially lead to very different outcomes.
One should also vary
protoplanet masses~$M_{\rm em}$,
their numbers, or
their initial spacing in terms of the mutual Hill radius,
because the dynamics is controlled
not only by properties of {\em individual\/} protoplanets
(disk-driven migration rates, etc.),
but also by mutual interactions within the whole system.
Consequently, the main goal of this paper is to study
the evolution of multiple planets embedded in gas/pebble disks
and its dependence on parameters.

Knowing the correct migration rates,
damping and pumping of $e$'s and $i$'s,
is also important for non-hydrodynamic models.
For example~\cite{Coleman_Nelson_2016MNRAS.457.2480C} or \cite{Izidoro_etal_2017MNRAS.470.1750I}
used an N-body model with parametrized migration rates
to explain configurations of compact Kepler planetary systems.
In their case all bodies of a given size drift in a systematic way,
because the disk torques were estimated for single planets,
and no heating torque was included.
Hereinafter, we shall see the situation is actually more complicated.

Our paper is organized as follows.
In Section~\ref{sec:model} we describe the radiation-hydrodynamical model
and its common parameters. In Section~\ref{sec:results} we present results
of 8~different simulations, including detailed views of protoplanet
encounters. Section~\ref{sec:conclusions} is devoted to conclusions.
A broader context of our work is discussed in Appendix~\ref{sec:context}.


\section{Model}\label{sec:model}

Our 2D numerical model is based on \textsc{Fargo} \citep{Masset_2000A&AS..141..165M},
and was described in detail in~\cite{Chrenko_etal_2017A&A...606A.114C}.%
\footnote{See also {\tt http://sirrah.troja.mff.cuni.cz/\char`~chrenko/}.}
Nevertheless, in order to present a self-contained paper
we recall the system of radiation--hydrodynamic equations
(i.e. the continuity, Navier--Stokes, energy, continuity of pebbles,
momentum of pebbles, equation of state, and gravity on protoplanets)
and our notation here:
\begin{equation}
{\pd\Sigma\over\pd t} + \vec v\cdot\nabla\Sigma = -\Sigma\nabla\cdot\vec v - \left({\pd\Sigma\over\pd t}\right)_{\!\rm acc},\label{eq:dSigma_dt}
\end{equation}
\begin{equation}
{\pd\vec v\over\pd t} + \vec v\cdot\nabla\vec v = -{1\over\Sigma}\nabla P + {1\over\Sigma}\nabla\!\cdot\!{\sf T} - {\int\rho\nabla\!\phi\,\d z\over\Sigma} + {\Sigma_{\rm p}\over\Sigma}{\Omega_{\rm K}\over\tau}(\vec u-\vec v)\,,
\end{equation}
\vskip-\baselineskip
\begin{eqnarray}
{\pd E\over\pd t} + \vec v\cdot\nabla E &=& -E\nabla\cdot\vec v - P\nabla\cdot\vec v + Q_{\rm visc} + {2\sigma T_{\rm irr}^4\over\tau_{\rm eff}} - {2\sigma T^4\over\tau_{\rm eff}}+ \nonumber\\
&& +\, {2H\nabla\cdot{{16\sigma\lambda_{\rm lim}\over\rho_0\kappa_{\rm R}}}T^3\nabla T} + \sum_i{GM_i\dot M_i\over R_i S_{\!\rm cell}}\delta(\vec r_i)\,,
\end{eqnarray}
\begin{equation}
{\pd\Sigma_{\rm p}\over\pd t} + \vec u\cdot\nabla\Sigma_{\rm p} = -\Sigma_{\rm p}\nabla\cdot\vec u - \left({\pd\Sigma_{\rm p}\over\pd t}\right)_{\!\rm acc},
\end{equation}
\begin{equation}
{\pd\vec u\over\pd t} + \vec u\cdot\nabla\vec u = -{\int\rho_{\rm p}\nabla\!\phi\,\d z\over\Sigma_{\rm p}} - {\Omega_{\rm K}\over\tau}(\vec u-\vec v)\,,
\end{equation}
\begin{equation}
P = \Sigma {RT\over\mu} = (\gamma-1)E\,,
\end{equation}
\vskip-\baselineskip
\begin{eqnarray}
\ddot{\vec r}_i &=& -{GM_\star\over r_i^3}\vec r_i - \sum_{j\ne i}{GM_j\over|\vec r_i-\vec r_j|^3}(\vec r_i-\vec r_j) \,- \nonumber\\
&& -\, \int\!\!\!\int{G\kern.5pt\Sigma\over|\vec r_i-\vec r_{\rm cell}|^3}(\vec r_i-\vec r_{\rm cell})r\d\theta\d r + f_z\kern1pt\hat z \quad\hbox{for }\forall i\,,\label{eq:ddot_r}
\end{eqnarray}
where
$\Sigma$~denotes the gas surface density,
$\vec v$~gas velocity,
$(\pd\Sigma/\pd t)_{\rm acc}$~gas accretion term,
$P$~vertically integrated pressure,
$\sf T$~viscous stress tensor,
$\rho$~gas volumetric density,
$\phi$~gravitational potential of the Sun and protoplanets,
with a cubic smoothing due to a finite cell size \citep{Klahr_Kley_2006A&A...445..747K},
$z$~vertical coordinate,
$\Sigma_{\rm p}$~pebble surface density,
$\vec u$~pebble velocity,
$(\pd\Sigma_{\rm p}/\pd t)_{\rm acc}$~pebble accretion term
for both Bondi and Hill regimes
(detailed in \citealt{Chrenko_etal_2017A&A...606A.114C}),
$\Omega_{\rm K}$~the Keplerian angular velocity,
$\tau$~the Stokes number of pebbles,
always assuming the Epstein drag regime,
$E$~gas internal energy,
$Q_{\rm visc}$~viscous heating term \citep{Mihalas_1984frh..book.....M},
$\sigma$~the Stefan--Boltzmann constant,
$T_{\rm irr}$~irradiation temperature \citep{Chiang_Goldreich_1997ApJ...490..368C},
$\tau_{\rm eff}$~effective optical depth \citep{Hubeny_1990ApJ...351..632H},
$T$~gas temperature,
$H$~vertical (pressure) scale height,
$\lambda_{\rm lim}$~flux limiter \citep{Kley_1989A&A...208...98K},
$\rho_0$~midplane density,
$\kappa_{\rm R}$~the Rosseland opacity
(the Planck opacity $\kappa_{\rm P}$ hidden in $\tau_{\rm eff}$ is assumed the same),
$G$~gravitational constant,
$M_i$~protoplanet mass,
$R_i$~protoplanet radius,
$S_{\!\rm cell}$~cell area in which it is located,
$\mu$~mean molecular weight,
$\gamma$~adiabatic index,
$\ddot{\vec r}_i$~gravitational acceleration of the body~$i$,
where a smoothing is applied again for the 3rd term;
and $f_z$~vertical damping prescription \citep{Tanaka_Ward_2004ApJ...602..388T}.

More specifically, we use the following smoothing of the potential:
\begin{equation}
\phi_i(d) = \cases{
-{GM_i\over d} & for $d > r_{\rm sm}$\,, \cr
-{GM_i\over d}\left[\left({d\over r_{\rm sm}}\right)^4 - 2\left({d\over r_{\rm sm}}\right)^3 + 2{d\over r_{\rm sm}}\right] & for $d \le r_{\rm sm}$\,,
}
\end{equation}
with the smoothing length $r_{\rm sm} = 0.5R_{\rm H}$,
where $R_{\rm H}$ denotes the Hill radius of the respective protoplanet.
We still integrate in the vertical direction over the density profile:
\begin{equation}
\rho(z) = {\Sigma\over\sqrt{2\pi}H}\exp\left(-{z^2\over2H^2}\right)
\end{equation}
to avoid a~smoothing over~$H$ \citep{Muller_etal_2012A&A...541A.123M}.

We performed a few modifications of the code, namely
we improved the successive over-relaxation (SOR) solver for the radiative step,
so that iterations stop when the system of equations is fulfilled at the machine precision.
We also included the gas accretion term using \cite{Kley_1999MNRAS.303..696K} prescription with $f_{\rm acc}$ parameter,
even though it is switched off in most simulations.
\cite{Bell_Lin_1994ApJ...427..987B} LTE opacities were implemented more carefully
without even minor jumps at the transitions.

The nominal simulation presented in~\cite{Chrenko_etal_2017A&A...606A.114C},
hereinafter called \verb|CaseIII_nominal|,
had the following parameters:
the gas surface density $\Sigma_0 = 750\,{\rm g}\,{\rm cm}^{-2}$ at 1\,au,
slope~$r^{-0.5}$,
the aspect ratio~$h = H/r$ and flaring are given by radiation processes during the relaxation phase
(one can start with an arbitrary value, $h = 0.02$ or 0.10, and the relaxation will converge to the same disk structure);
adiabatic index $\gamma = 1.4$,
molecular weight $\mu = 2.4\,{\rm g}\,{\rm mol}^{-1}$,
vertical opacity drop $c_\kappa = 0.6$,
effective temperature $T_\star = 4370\,{\rm K}$,
stellar radius $R_\star = 1.5\,R_\odot$,
disc albedo $A = 0.5$,
softening parameter is~$0.5R_{\rm H}$,
the entire Hill sphere is considered when calculating disk$\,\leftrightarrow\,$planet interactions,
the inner boundary is $r_{\rm min} = 2.8\,{\rm au}$,
outer boundary $r_{\rm max} = 14\,{\rm au}$,
a~damping BC (e.g. \citealt{Kley_Dirksen_2006A&A...447..369K}) is used to prevent spurious reflections,
and applied up to $1.2r_{\rm min}$ and from $0.9r_{\rm max}$ on;
quantities are damped towards their initial values,
only during the relaxation phase the damping is towards zero radial velocity;
the vertical damping parameter is~0.3,
pebble flux $\dot M_{\rm p} = 2\times10^{-4}\,M_\oplus\,{\rm yr}^{-1}$,
turbulent stirring parameter $\alpha_{\rm p} = 10^{-4}$,
the Schmidt number ${\rm Sc} = 1$,
pebble coagulation efficiency $\epsilon_{\rm p} = 0.5$,
pebble bulk density $\rho_{\rm p} = 1\,{\rm g}\,{\rm cm}^{-3}$,
embryo density $\rho_{\rm em} = 3\,{\rm g}\,{\rm cm}^{-3}$ (constant),
embryo mass $M_{\rm em} = 3.0\,M_\oplus$,
and their number $N_{\rm em} = 4$.


For simplicity, we used a constant kinematic viscosity
$\nu = 5.0\times10^{14}\,{\rm cm}^2\,{\rm s}^{-1} \doteq 10^{-5}$ [c.u.],
but we do not expect a drastic change when we would use an $\alpha$-viscosity,
with $\nu = \alpha c_{\rm s} H$,
where $c_{\rm s}$ denotes the (local) sound speed
and $H$ the scale height
\citep{Shakura_Sunyaev_1973A&A....24..337S}.
The viscously-heated region is actually from 3~to 7 au in our nominal model,
the rest is irradiated. We also do not expect (prescribe) any jumps in viscosity
due to MRI/dead regions, as our disk is cold throughout.

There are several limitations of our model we shall keep in mind.
We do not account for the Stokes drag
(i.e. pebble sizes $D_{\rm p}$ larger than the mean-free path~$\ell$),
a reduction factor of the accretion rate needed for $\tau \gg 1$
\citep{Ormel_Klahr_2010A&A...520A..43O,Ida_etal_2016A&A...591A..72I},
or the gravity of pebbles
(as emphasized by \citealt{Benitez_Pessah_2018ApJ...855L..28B}).
Nevertheless, our gas/pebble disks should be well within the respective limits.
We find our typical pebble-to-gas ratio to be approximately 0.001,
and if assume for the moment that the results of \cite{Benitez_Pessah_2018ApJ...855L..28B}
are applicable even in our more complicated case
(i.e. non-fixed planets, eccentric orbits, with pebble accretion, back-reaction),
the ratio of torques should be
$\Gamma_{\rm p}/\Gamma_{\rm g} \doteq 0.05$.
The pebble torque can thus be safely ignored for our disc conditions.

The discretisation in space we normally use is $1024\times 1536$ cells
in the radial and azimuthal directions.
See Appendix~\ref{sec:convergence} for an additional convergence test.
The discretisation in time is controlled by the CFL condition;
the maximal time step is $\Delta t = 3.725\,[{\rm c.u.}] = 1/20\,P_{\rm orb}$ at 5.2\,au.
Orbital elements are output every $20\,\Delta t$,
and hydrodynamical fields every $500\,\Delta t$.
The nominal time span is approximately 50\,kyr,
but the simulations are prolonged whenever needed.
The relative precision of the IAS15 integrator \citep{Rein_Spiegel_2015MNRAS.446.1424R}
used for the planetary bodies is set to $10^{-9}$.


\section{Results}\label{sec:results}

Apart from the nominal case, we performed 8~simulations
which are summarised in Table~\ref{sim_tab}.
We always change one parameter (or two at most)
and analyze how the overall evolution changes.
The dependence on
the gas surface density~$\Sigma$,
pebble flux~$\dot M_{\rm p}$,
viscosity~$\nu$,
embryo mass~$M_{\rm em}$,
or their number~$N_{\rm em}$
is a very basic question, indeed.
We thus shall describe all of them (not-so-interesting included),
with the most interesting implications discussed later in Section~\ref{sec:conclusions}.

A very useful tool would be a construction of a complete Type-I migration map,
i.e. a dependence of the torque~$\Gamma$ on the disk profiles $\Sigma(r)$, $T(r)$,
the protoplanet mass~$M_{\rm em}$, and other parameters,
in a similar way as in \cite{Paardekooper_etal_2011MNRAS.410..293P,Bitsch_etal_2013A&A...549A.124B}.
In our case we have an additional parameter, namely the pebble flux $\dot M_{\rm p}$,
which makes this task more difficult though.
It may also vary with the Stokes number~$\tau$ (cf.~\citealt{Benitez_Pessah_2018ApJ...855L..28B}).
At the same time, we would need an eccentricity excitation map,
for the derivatives~$\dot e$, and also for the asymptotic values~$e_{\rm asy}$.
As we shall see in the next Sections, the situation is even worse,
because there are mutual (hydrodynamical) interactions
between the protoplanets too.

\begin{table}
\caption{Selected parameters of our hydrodynamical simulations, where
$\Sigma_0$~denotes the gas surface density at 1\,au,
$\dot M_{\rm p}$~the pebble flux,
$\nu$~the kinematic viscosity,
$M_{\rm em}$~the embryos' mass, and
$N_{\rm em}$~their number.
For other parameters see the main text.}
\centering
\begin{tabular}{ll}
\hline
\vline height 8pt width 0pt
simulation & parameters \\
\hline
\vline height 9pt width 0pt
\verb|CaseIII_nominal|		& $\Sigma_0 = 750\,{\rm g}\,{\rm cm}^{-2}$ at 1\,au, \\
				& $\dot M_{\rm p} = 2\times10^{-4}\,M_\oplus\,{\rm yr}^{-1}$, \\
				& $\nu = 5.0\times10^{14}\,{\rm cm}^2\,{\rm s}^{-1} \doteq 10^{-5}$ [c.u.], \\
				& $M_{\rm em} = 3.0\,M_\oplus$, $N_{\rm em} = 4$, \\
\vline width 0pt depth 4pt
				& $f_{\rm acc} = 0$ \\
\hline
\vline height 10pt width 0pt
\verb|Sigma_3times|		& $\Sigma_0 = 2250\,{\rm g}\,{\rm cm}^{-2}$ \\
\verb|Sigma_1over3|		& $\Sigma_0 = 250\,{\rm g}\,{\rm cm}^{-2}$ \\
\verb|pebbleflux_2e-5|		& $\dot M_{\rm p} = 2\times10^{-5}\,M_\oplus\,{\rm yr}^{-1}$ \\
\verb|viscosity_1e-6|		& $\nu = 5.0\times10^{13}\,{\rm cm}^2\,{\rm s}^{-1} \doteq 10^{-6}$ [c.u.] \\
\verb|gasaccretion_1e-6|	& $f_{\rm acc} = 10^{-6}$ \\
\verb|totmass_20ME|		& $M_{\rm em} = 5.0\,M_\oplus$, $\dot M_{\rm p} = 2\times10^{-5}\,M_\oplus\,{\rm yr}^{-1}$ \\
\verb|embryos_1.5ME_8|		& $M_{\rm em} = 1.5\,M_\oplus$, $N_{\rm em} = 8$ \\
\vline width 0pt depth 4pt
\verb|embryos_0.1ME_120|	& $M_{\rm em} = 0.1\,M_\oplus$, $N_{\rm em} = 120$ \\
\hline
\end{tabular}
\label{sim_tab}
\end{table}


\begin{figure}
\centering
\includegraphics[width=8.5cm]{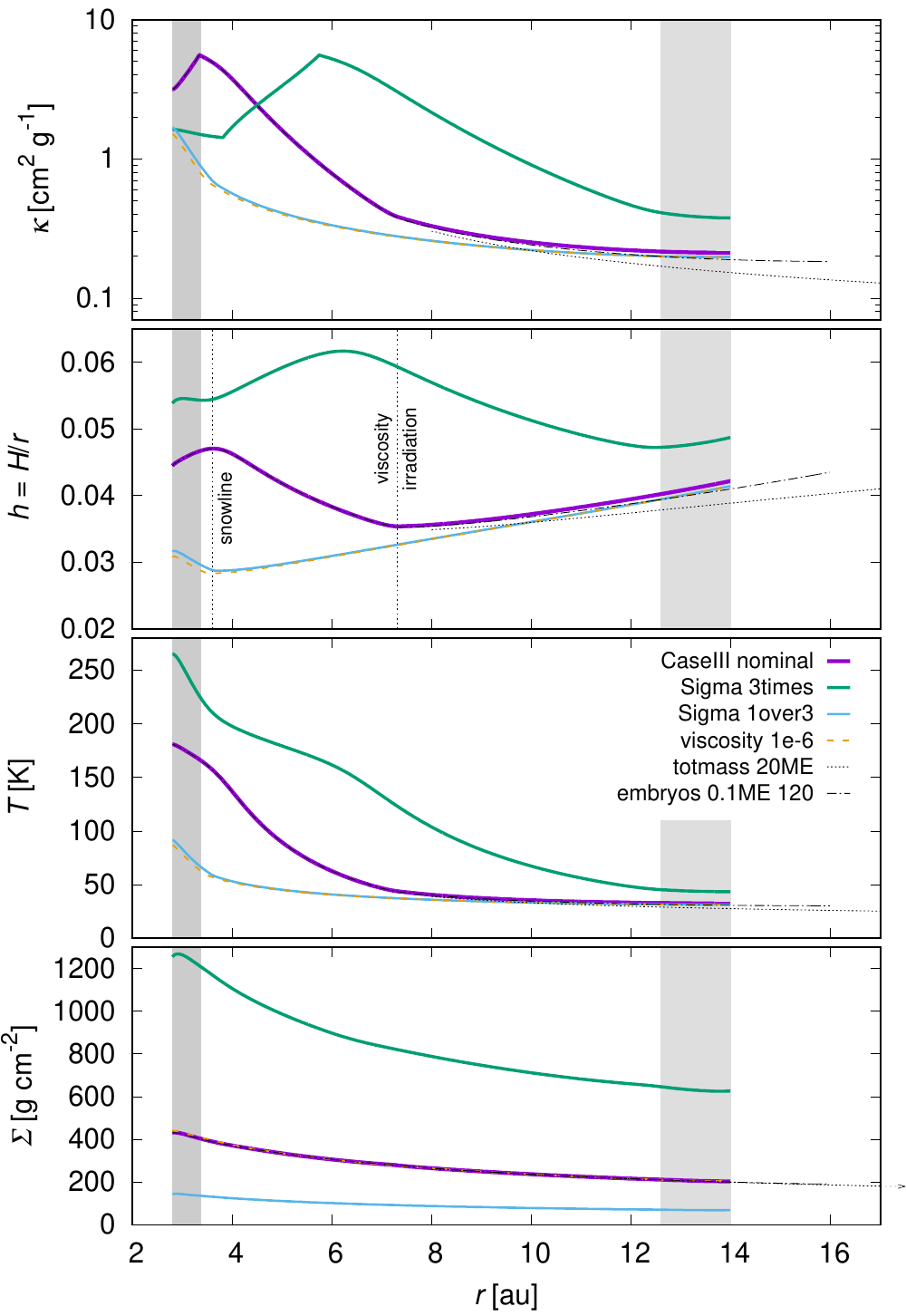}
\caption{Profiles of the gas disk after the relaxation phase,
which serve as the initial conditions for our simulations.
From top to bottom:
the opacity~$\kappa$,
the aspect ration~$h = H/r$,
the temperature~$T$,
and the surface density~$\Sigma$.
The gray boxes indicate the extent of the damping zones
at the inner and outer BC.
The vertical dotted lines show the snowline, and the transition
between the viscous heating and the stellar irradiation regions,
in case of the nominal simulation.
Some of the simulations actually use the same disk as the nominal one.
For {\tt totmass\char`_20ME} simulation we had to use
an outer disk, spanning from 8~to 40\,au.
Similarly, we used a slightly larger disk (up to 16\,au) for
the simulation {\tt embryos\char`_0.1ME\char`_120}.}
\label{profiles_kappa}
\end{figure}

\begin{figure}
\centering
\includegraphics[width=8.5cm]{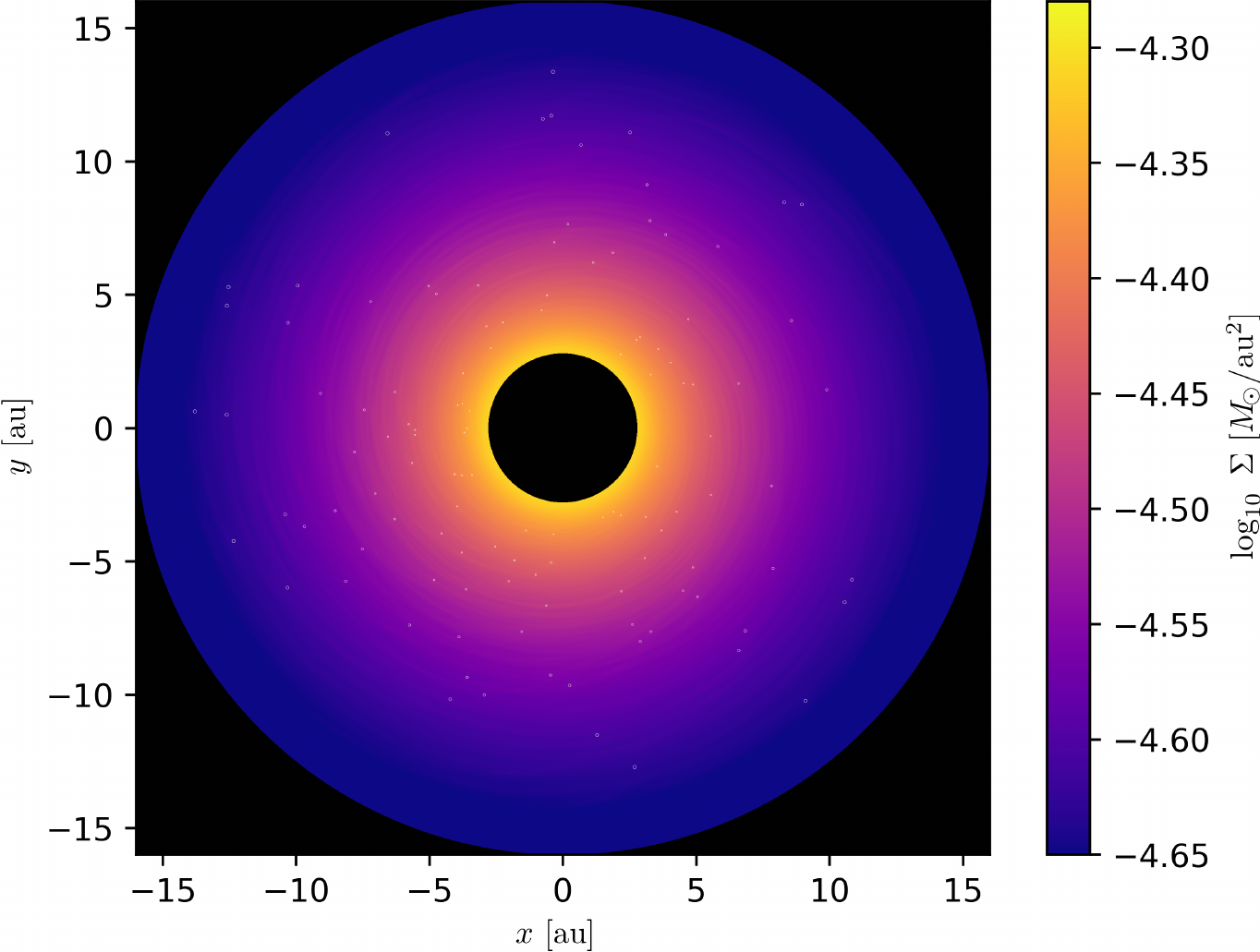}
\includegraphics[width=8.5cm]{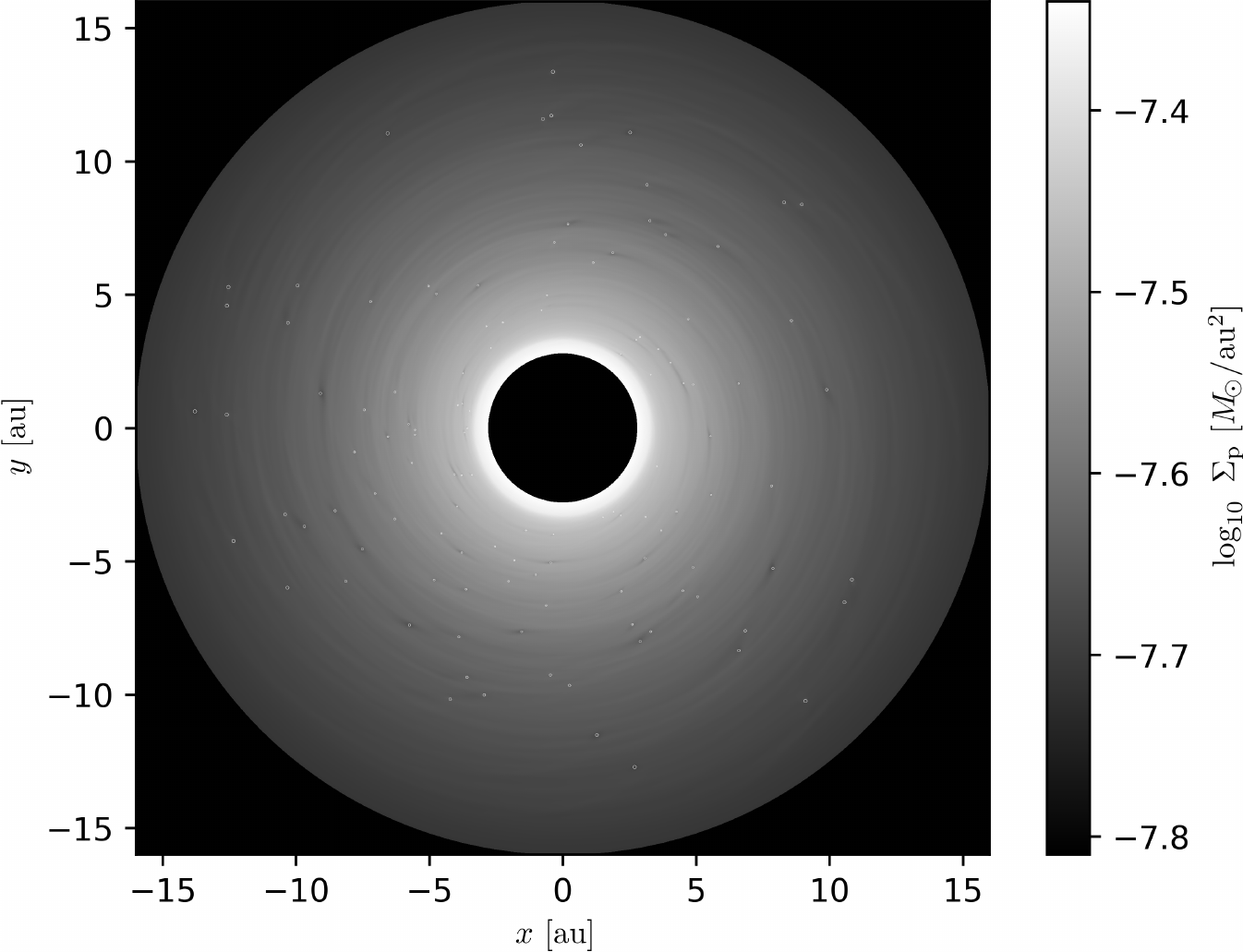}
\caption{An initial global structure of the gas and pebble disks
in one of our simulations. The gas surface density~$\Sigma$ (top)
and the pebble surface density~$\Sigma_{\rm p}$ (bottom) are shown.
In our model the disks interact mutually by means of an aerodynamic drag,
and also with protoplanets by means of gravitation (in case of gas)
and accretion.}
\label{gasdens4_2}
\end{figure}

\subsection{Initial profiles}

The initial radial profiles of the gas disk are shown in Figure~\ref{profiles_kappa}.
They were obtained by a~relaxation procedure, prior to the simulation itself.
The initial azimuthal profiles are uniform.
Several simulations actually use the same gas disk as the nominal case,
namely \verb|pebbleflux_2e-5|, \verb|embryos_1.5ME_8|, and \verb|gasaccretion_1e-6|.
Another two are only extensions of the nominal case towards larger radii,
namely \verb|embryos_0.1ME_120| to 16\,au,
and \verb|totmass_20ME| from 8~to 40\,au, because it turned out
the convergence zone is located in the outer disk.
The profiles of the pebble disk are shown in Figure~\ref{profiles_pebble}
for comparison. We assume the corresponding Stokes numbers~$\tau$
are drift-limited and that initially the pebble flux~$\dot M_{\rm p}$
is independent of the radial distance.

Classically, we would expect a convergence zone somewhere between
the region driven by viscous heating, and irradiation (flaring).
This is no longer true when we include the pebble accretion heating too.
Coincidentally, profiles for \verb|Sigma_1over3|, and \verb|viscosity_1e-6|
are very similar, but the disk dynamics is naturally different.

A global structure of the gas and pebble disks is shown in Figure~\ref{gasdens4_2}.
Initially, they are very smooth but after mere hundred orbital periods
spiral arms in the gas disk are developed,
surroundings of each protoplanet and its corotation region is affected by the accretion heating,
and also accretion-related structures occur in the pebble disk.


\subsection{Nominal case}

\begin{figure}
\centering
\verb|CaseIII_nominal|\break
\includegraphics[width=9cm]{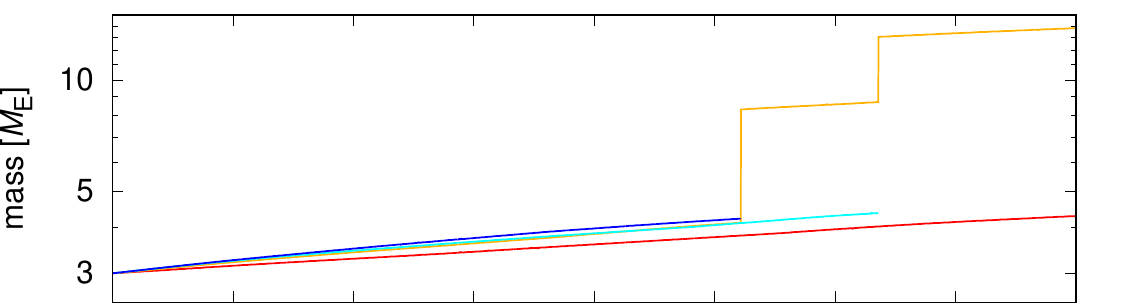}
\includegraphics[width=9cm]{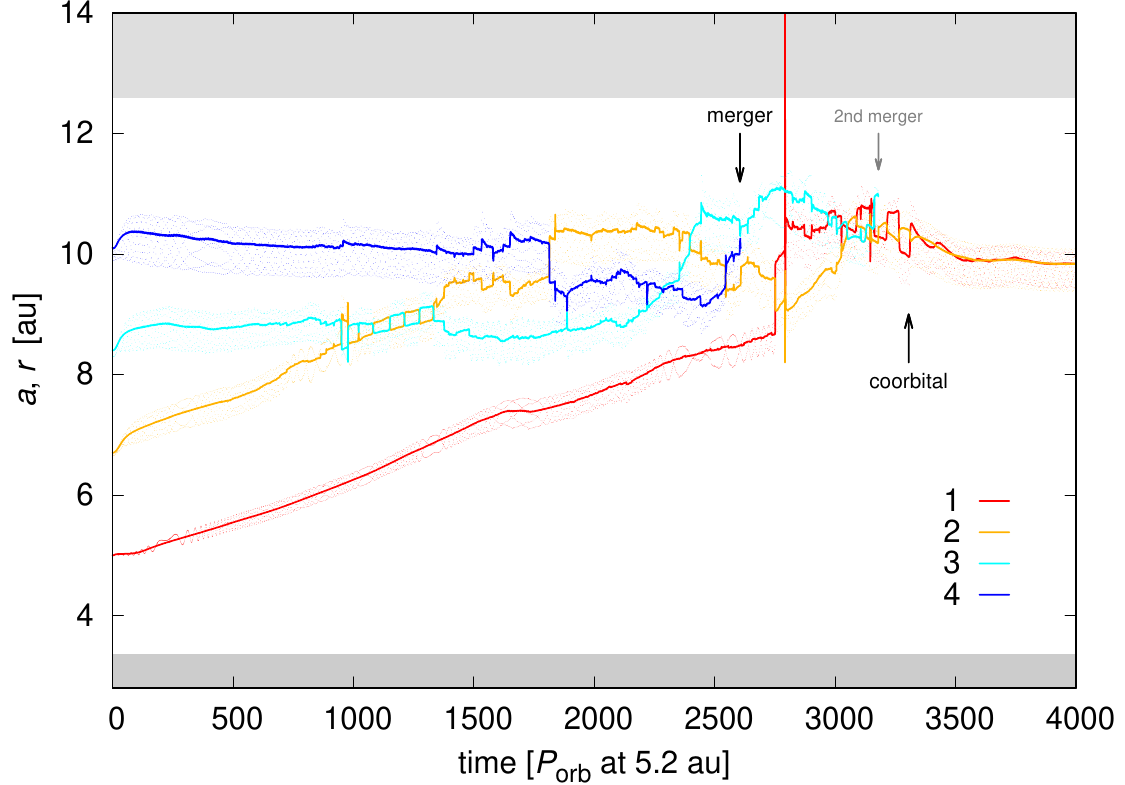}
\caption{The semimajor axis~$a$ (line), and the heliocentric distance~$r$ (dots) vs time~$t$ (bottom),
and the embryo mass~$M_{\rm em}$ vs~$t$ (top) for the nominal simulation with initially 4~embryos.
The time span $4000\,P_{\rm orb}$ corresponds to 47.4\,kyr.
Although it was already presented in \cite{Chrenko_etal_2017A&A...606A.114C},
we show it here to provide an easy (1:1) comparison.
The gray strips indicate the inner and outer damping regions
used to {\em gradually\/} suppress spurious reflections.
The black arrows indicate two interesting events we study in detail:
a~merger (3+4), and a~coorbital formation.
The semimajor axis may exhibit a `spike' during a~close encounter,
but in fact $r \ll a$; the body is far from the damping zone.}
\label{nbody.orbits.at0}
\end{figure}

Let us start with a~{\bf warning}: for substantially non-Keplerian orbits,
it is rather important to plot the radial distance~$r$
instead of~$a, q, Q$ (the semimajor axis, pericentre, and apocentre).
For example a~spiral orbit has a~perfectly smooth $r(t)$,
with no oscillations whatsoever, but its eccentricity $e \gg 0$.
One could be misled by non-zero $e$ and think of close encounters between such orbits,
but in fact there is always a substantial separation
(cf.~Fig.~\ref{nbody.orbits.at0}).

We briefly recall the hot-trail effect is visible soon after the beginning of the simulation;
at $t \simeq 100\,P_{\rm orb}$ there are developed oscillations of~$r(t)$
(which would correspond to the eccentricity up to~$0.035$).
The zero-torque radius in the absence of accretion heating would be
located at about 7\,au where the disk becomes flared by stellar irradiation.
Instead, we see that planets migrate towards 9\,au. Such an offset
is due to the accretion heating, which adds a positive torque contribution,
and also the hot-trail effect, because the Lindblad and corotation
torques become modified for oscillating orbits.

No low-order mean-motion resonances (MMR) are established during migration,
partly because embryos were initially too close,
so no captures are expected even if the simulation is run longer.
Resonances 3:2, 4:3, 5:4, \dots were encountered,
but a resonant capture would be difficult anyway,
because $e > 0$ \citep{Batygin_2015MNRAS.451.2589B}.
During a series of close encounters, there are about
20~exchanges when orbits radially swap,
5~repulsions when the distance between orbits increases, and
2~successful mergers, $13.8$ and $4.3\,M_\oplus$,
which finally settle to a coorbital configuration (Figure~\ref{nbody.orbits.at0}).

In the following, we show details of two representative events:
the merger and the coorbital formation. As a novelty, we realized
that 3-body interactions are needed for successful mergers!
One can see this already from $r(t)$, that a third embryo
is `always' (2 out of 2 cases) present in the vicinity.

\newdimen\tmp\tmp=3.7cm
\begin{figure*}
\centering
\includegraphics[height=\tmp]{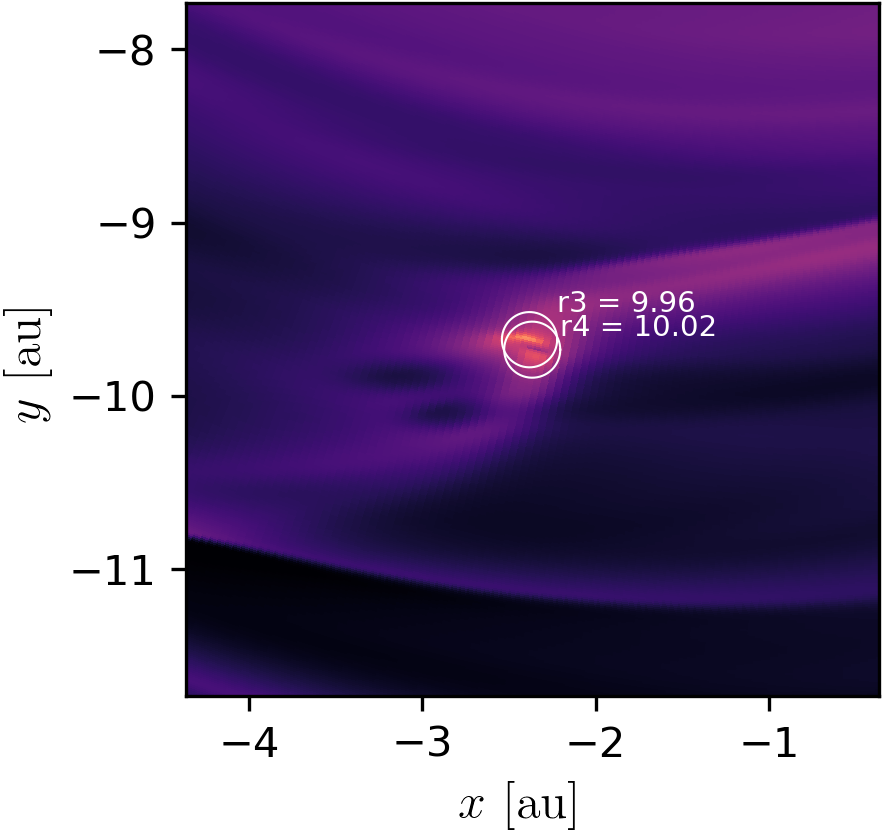}
\includegraphics[height=\tmp]{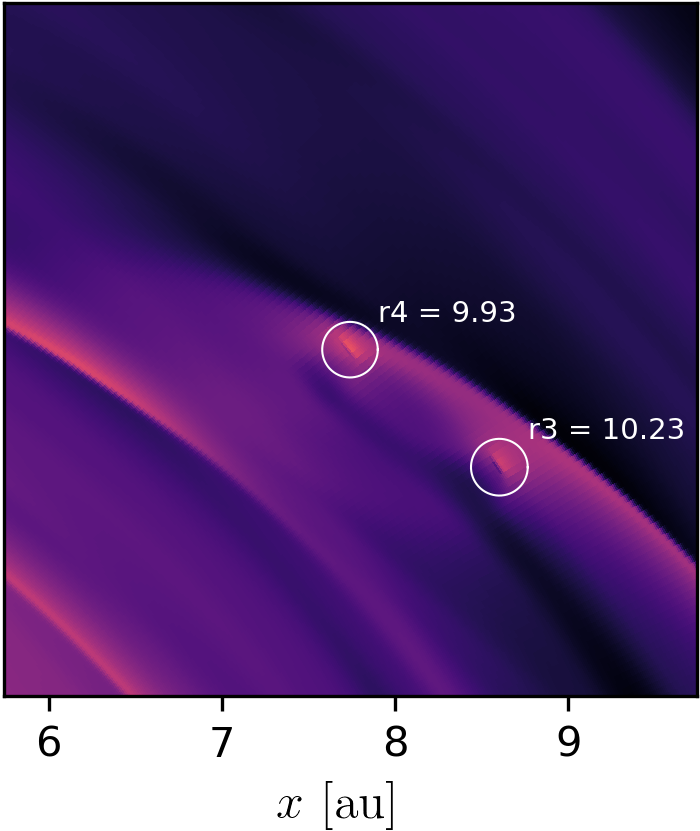}
\includegraphics[height=\tmp]{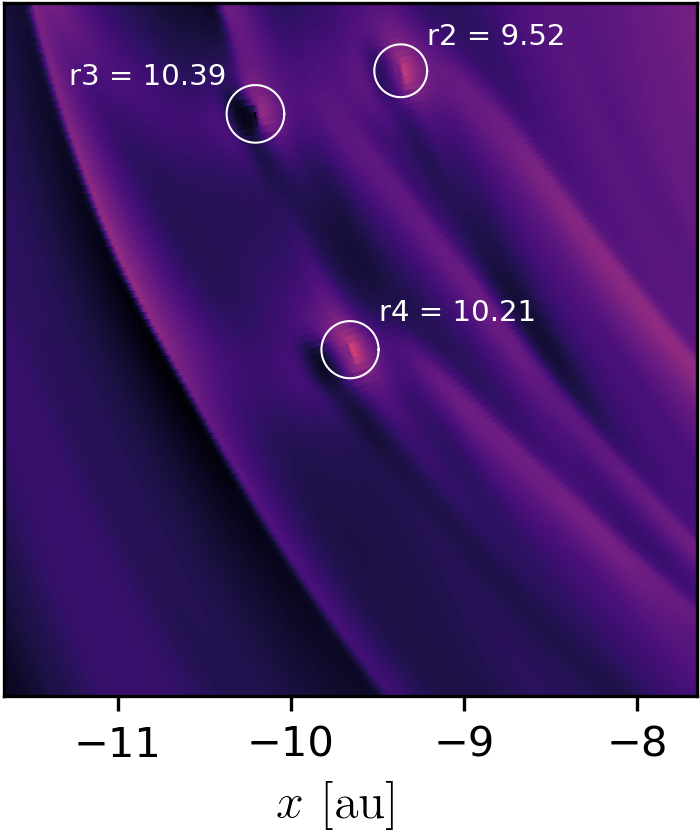}
\includegraphics[height=\tmp]{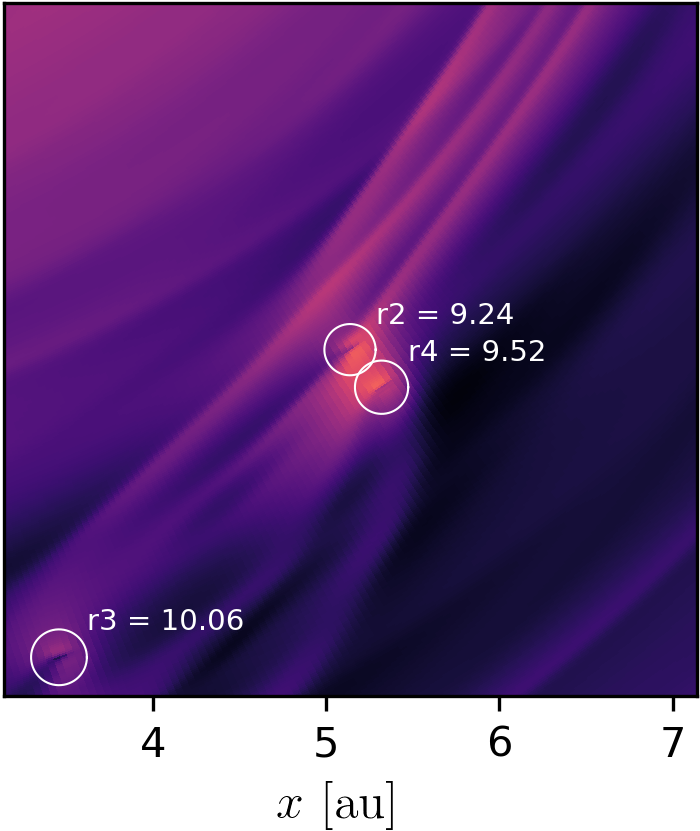}
\includegraphics[height=\tmp]{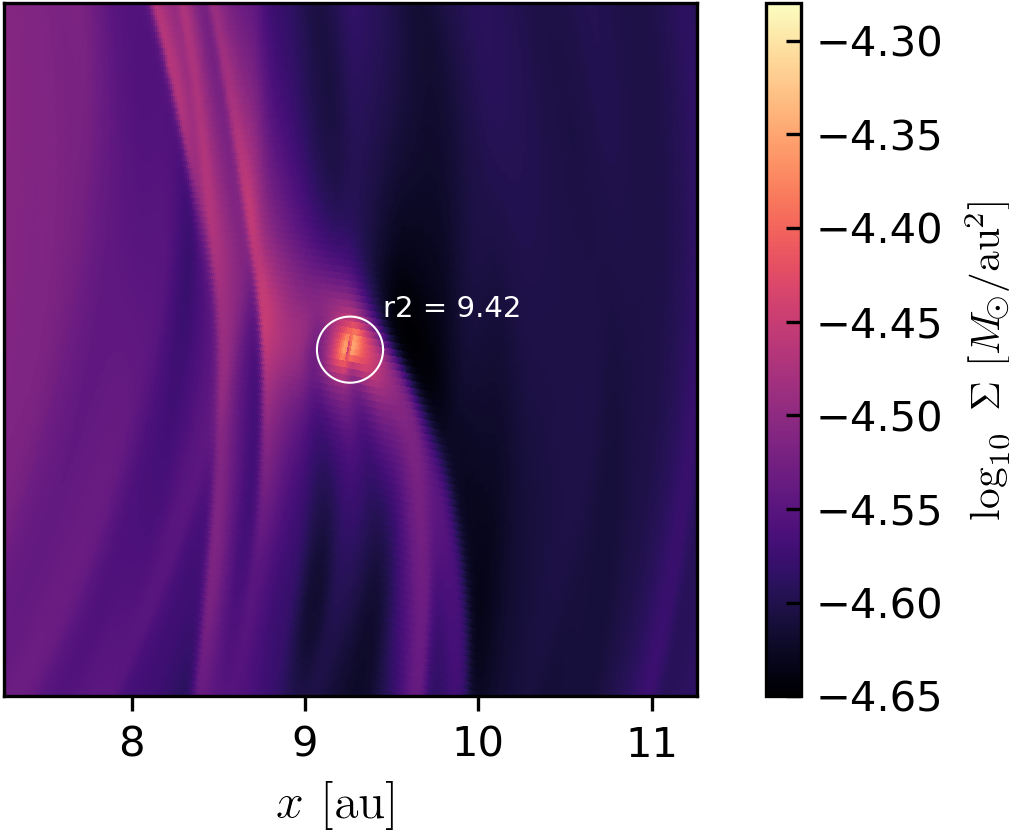}
\caption{The gas surface density $\Sigma$ in the $(x,y)$ plane
for a very short segment ($t = 2605$ to $2610\,P_{\rm orb}$ at 5.2\,au)
of the {\tt CaseIII\char`_nominal} simulation, showing approximately 1~long-period orbit
during which a 3-body interaction occurs, and results in a merger event.
The $y$-coordinates are different for individual panels,
but the overall range is always $4\,{\rm au}$.
The Sun is located at $(0,0)$.
The positions of the embryos are indicated by their Hill spheres (circles),
and heliocentric distances (labels, in ${\rm au}$ units).
The motion is in a counter-clockwise sense.
}
\label{CaseIII_nominal_Z1_MERGER1}
\end{figure*}

\begin{figure}
\centering
\includegraphics[width=8cm]{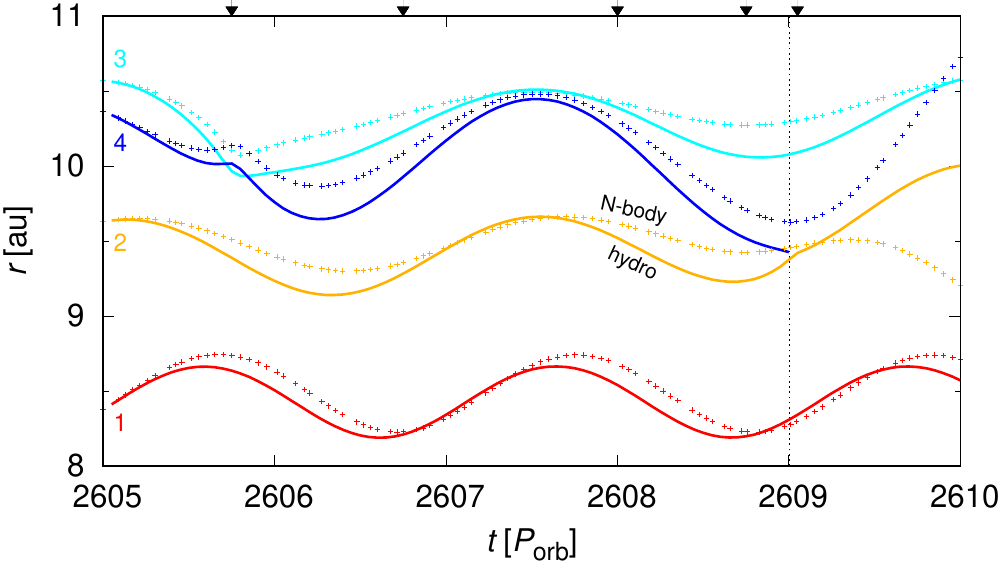}
\caption{The heliocentric distance~$r$ vs time~$t$ for all four embryos
in the simulation shown in Fig.~\ref{CaseIII_nominal_Z1_MERGER1}.
The black triangles correspond to the time of individual snapshots.
There is an output from the full hydrodynamic simulation (lines),
and an N-body simulation with no disks, no torques (points) for comparison.
The latter was restarted from the very same initial conditions, at
$t = 2605\,P_{\rm orb}$.
One can see the merger event at
$t \doteq 2609\,P_{\rm orb}$.
In case of the N-body simulation, the evolution is different,
because without the disk torques the trajectories are mostly Keplerian,
the encounter between embryos 3\,and\,4 one orbit prior to the merger (2+4)
has a different geometry, so the merger actually does not occur.}
\label{CaseIII_nominal_Z1_MERGER1_rt}
\end{figure}

\begin{figure}
\centering
\includegraphics[width=8cm]{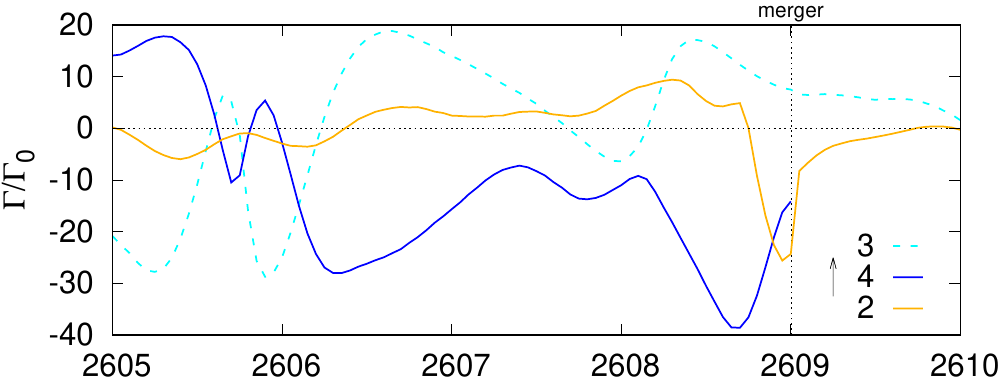}
\includegraphics[width=8cm]{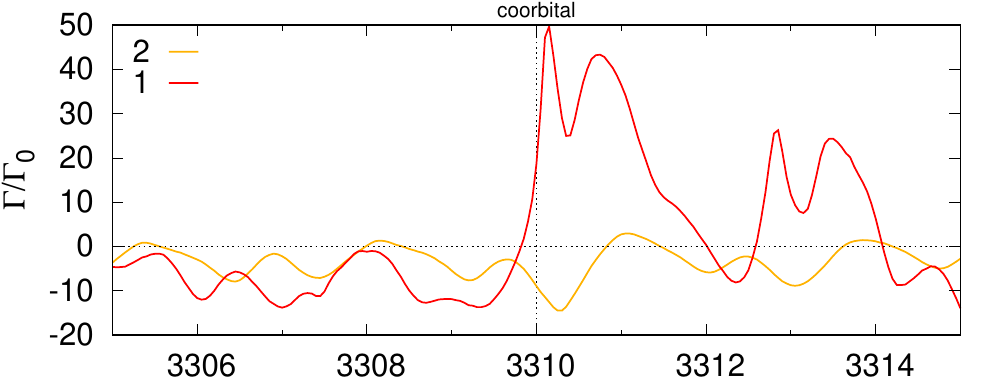}
\includegraphics[width=8cm]{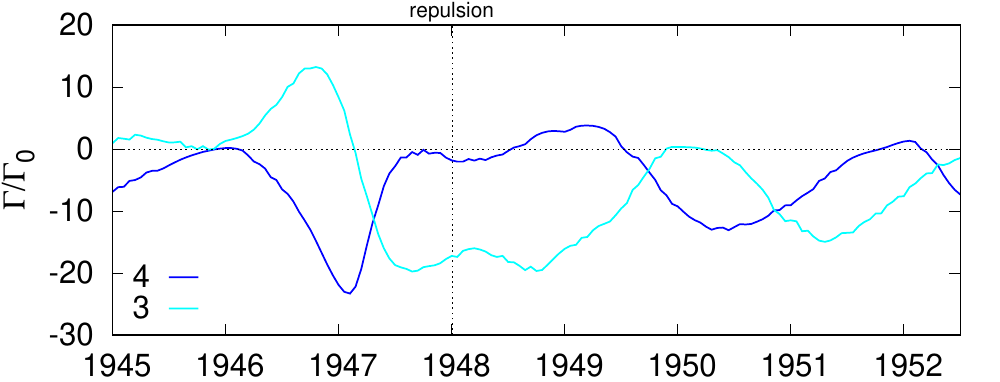}
\includegraphics[width=8cm]{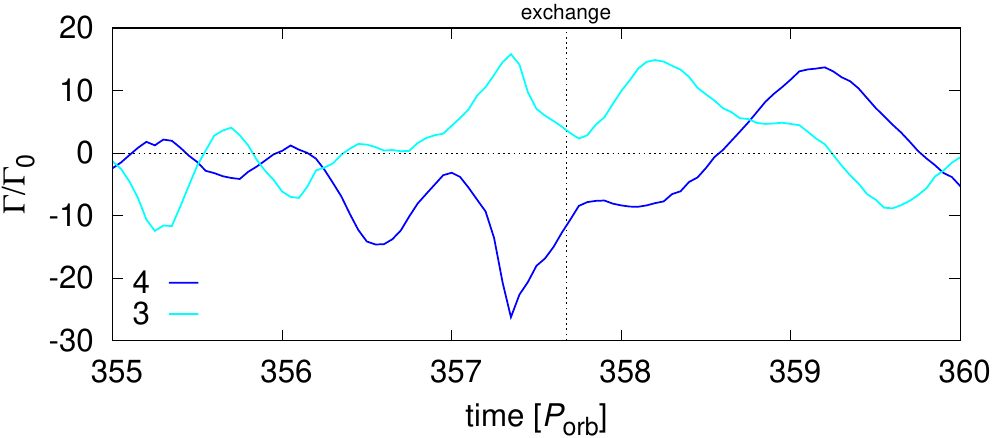}
\caption{The normalized total disk torque $\Gamma/\Gamma_0$ vs time~$t$,
where $\Gamma_0 = (q/h)^2\Sigma r^4 \Omega^2$, $q = M_{\rm em}/M_\star$,
for four events studied in detail:
a~merger (top),
a~coorbital formation,
a~repulsion, and
an~exchange (bottom).
These events were selected from the following simulations:
{\tt CaseIII\char`_nominal},
ditto,
{\tt Sigma\char`_3times},
{\tt viscosity\char`_1e-6}.
Only those embryos which take part in the interaction are plotted.
The label order corresponds to the (initial) radial distance.
Each case exhibits a~very different progression of the torque.
In particular, there is:
(i)~strong negative torque acting on the outer embryo (for the merger);
(ii)~positive torque on the inner embryo (coorbital);
(iii)~alternating torques on both (repulsion); and
(iv)~negative torque on the outer and positive on the inner (exchange).
}
\label{tqwk}
\end{figure}

\subsubsection{Detail: Merger}

In order to resolve a detail, the simulation has to be restarted
from hydrodynamical field files, prior to the time of interest.
We use at least 100 times finer output (every $\Delta t$).
We have to admit, that the evolution is not always {\em exactly\/} the same;
for example the SOR method (its relaxation factor) depends on past evolution.
Nevertheless, the merger event at $t \doteq 2609\,P_{\rm orb}$
was repeated perfectly, as shown in Figures~\ref{CaseIII_nominal_Z1_MERGER1},
\ref{CaseIII_nominal_Z1_MERGER1_rt}.
and~\ref{CaseIII_nominal_Z1_MERGER1_xyz}.

First, embryo~4 scatters off embryo~3 during a deep encounter,
with a minimum separation being a small fraction of the Hill radius,
$R_{\rm H} \doteq 0.15\,{\rm au}$.
Second, embryo~4 encounters embryo~2 during the {\em next\/} orbit, so they merge.
Without the prior strong perturbation, the collision would not occur.
At the same time, there are disk torques which substantially
affect the evolution (Figure~\ref{tqwk}).
Prior to the merger, they brake the outer embryo~4.
It seems that the relative motion in the $z$-direction
is not important in our situation, being a small fraction of the embryo radius,
$\Delta z \ll R_{\rm em} \simeq 10^{-4}\,{\rm au}$.

We also performed a test with a purely N-body integration,
in order to check, whether this event is caused either
by a mutual gravitational interaction between the embryos,
or by hydrodynamics. The integration was restarted from
the same time, the derivatives being the same, but without
any disk torques, the orbits are Keplerian most of the time,
so both encounters have different geometries and a merger event
does not occur at all. One can see the actual difference
in Fig.~\ref{CaseIII_nominal_Z1_MERGER1_rt}.


\subsubsection{Detail: Coorbital}

\begin{figure*}
\centering
\includegraphics[height=\tmp]{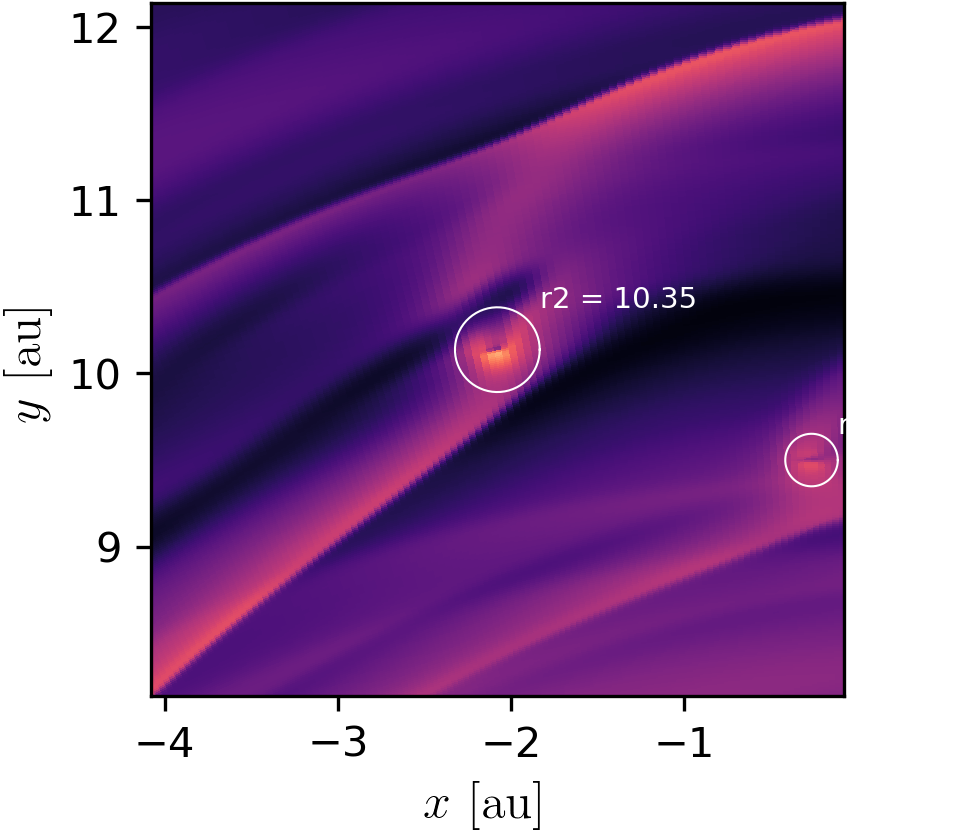}\kern-.5cm
\includegraphics[height=\tmp]{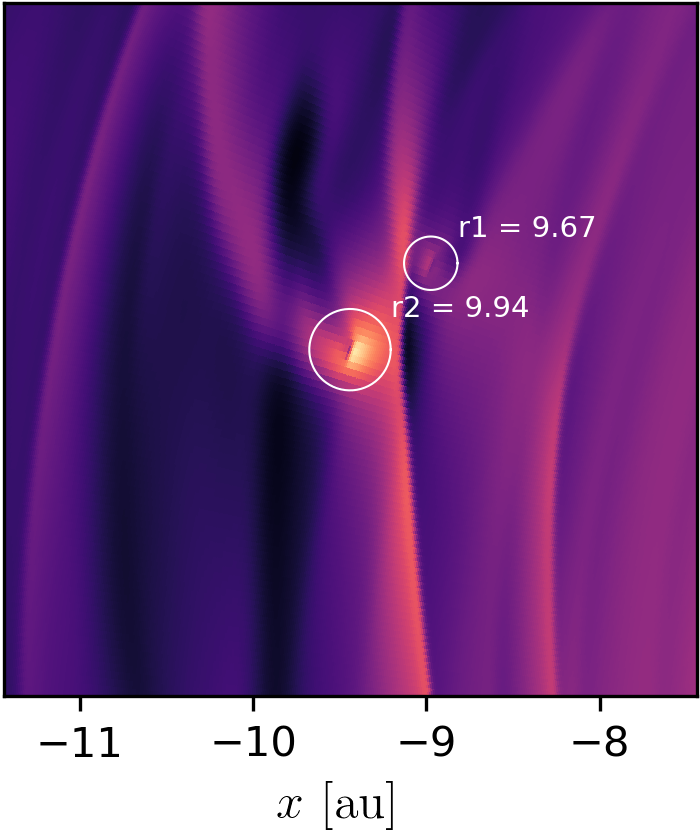}
\includegraphics[height=\tmp]{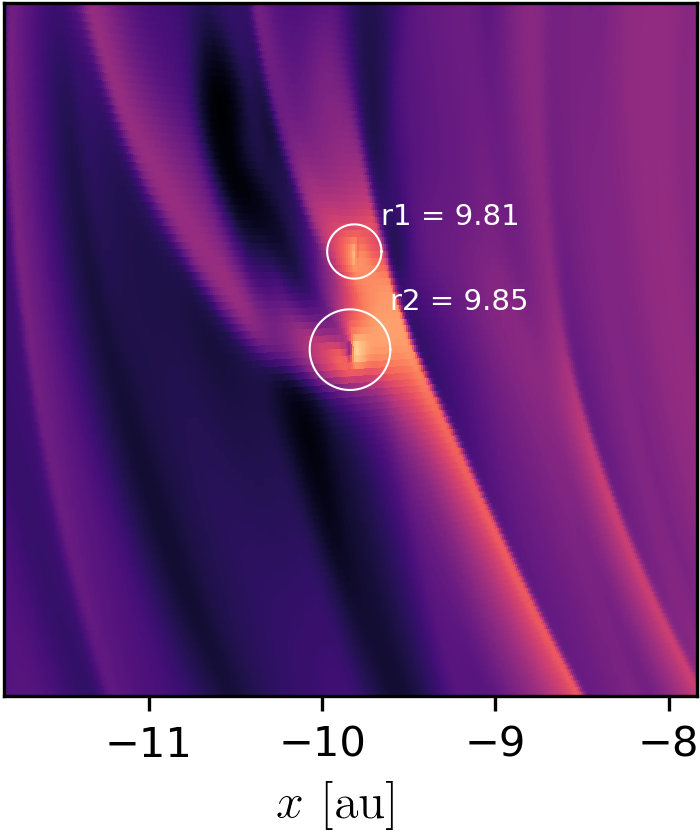}
\includegraphics[height=\tmp]{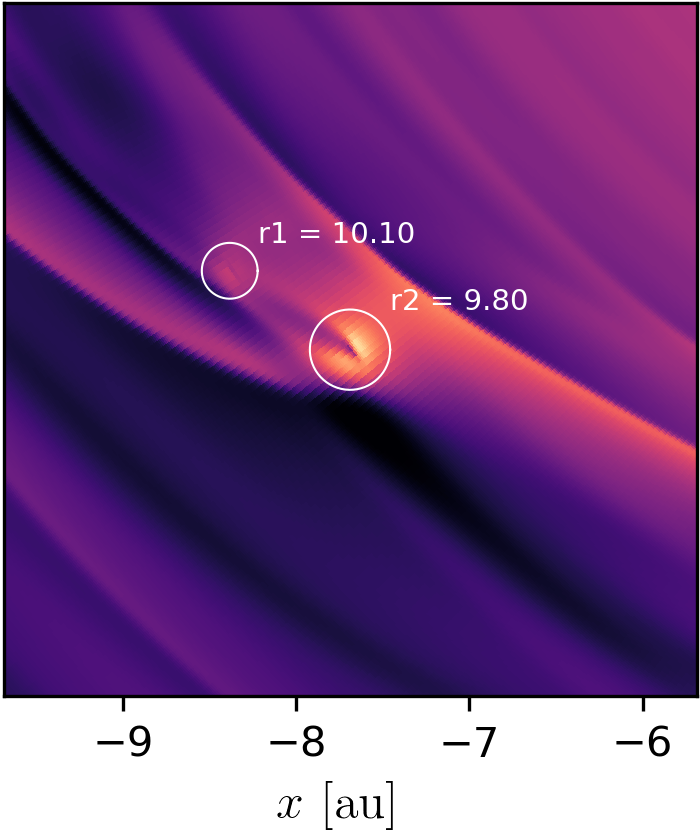}
\includegraphics[height=\tmp]{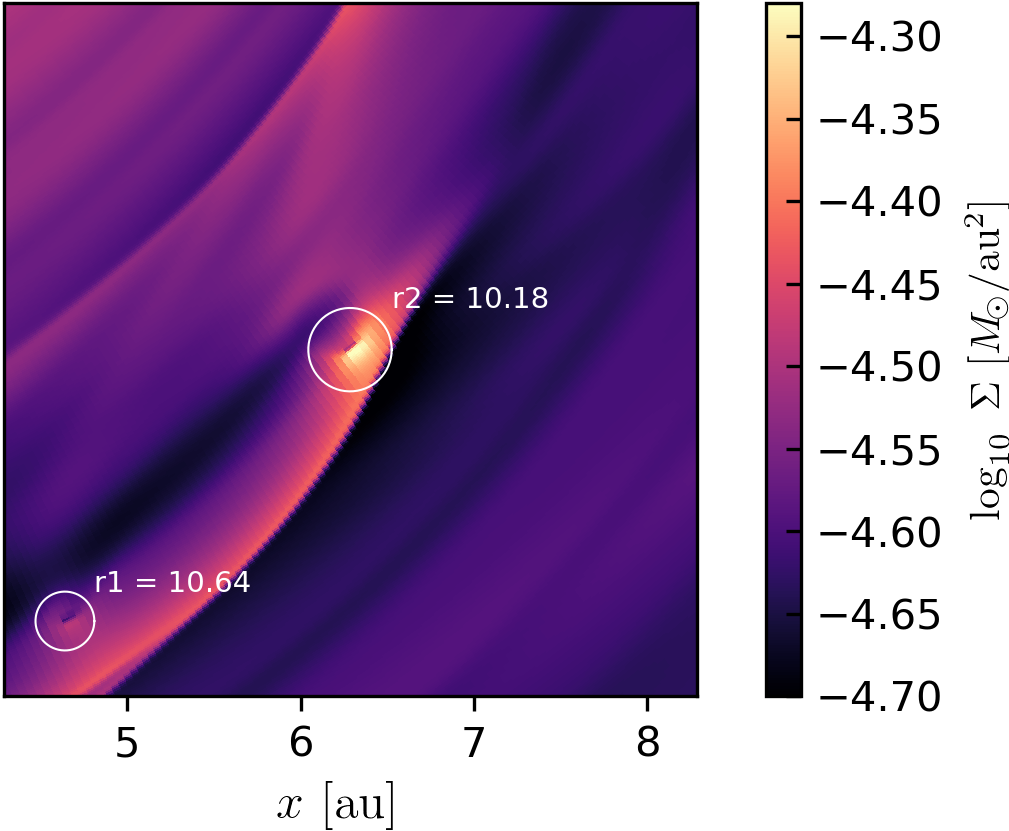}
\caption{The same as Fig.~\ref{CaseIII_nominal_Z1_MERGER1},
but for a coorbital formation during the interval $t = 3305$ to $3315\,P_{\rm orb}$.
The less-massive embryo approaches the more-massive from behind,
flying through its inner (detached) spiral arm,
enters the coorbital region,
and flies away in a low density region.
}
\label{CaseIII_nominal_Z3_COORBITAL}
\end{figure*}

As shown in Figure~\ref{CaseIII_nominal_Z3_COORBITAL},
the coorbital was formed at $t \doteq 3310\,P_{\rm orb}$
by an encounter of the less-massive embryo~1 with the more-massive embryo~2
from behind. During the approach, embryo~1 flies through a detached,
or prolonged spiral arm of embryo~2. The corresponding disk torque
suddenly changes from negative to positive (normalized $\Gamma \simeq 50$; Fig.~\ref{tqwk}).
We stress these torques would be calculated wrong in N-body models
which use prescriptions derived for single planets.
On the departure, the former embryo enters an underdense region
and is captured in the coorbital region of the latter.

While we do not model a long-term evolution of the coorbital pair here,
we can assume its stabilisation according to Sec.~\ref{sec:viscosity_1e-6}.
Similarly as before, without the disk torques, only an orbital exchange would occur.
Coorbitals are generally common outcomes of the simulations,
because the resonant captures are prevented by non-zero eccentricities.

\subsection{High surface density}

\begin{figure}
\centering
\verb|Sigma_3times|\break
\includegraphics[width=9cm]{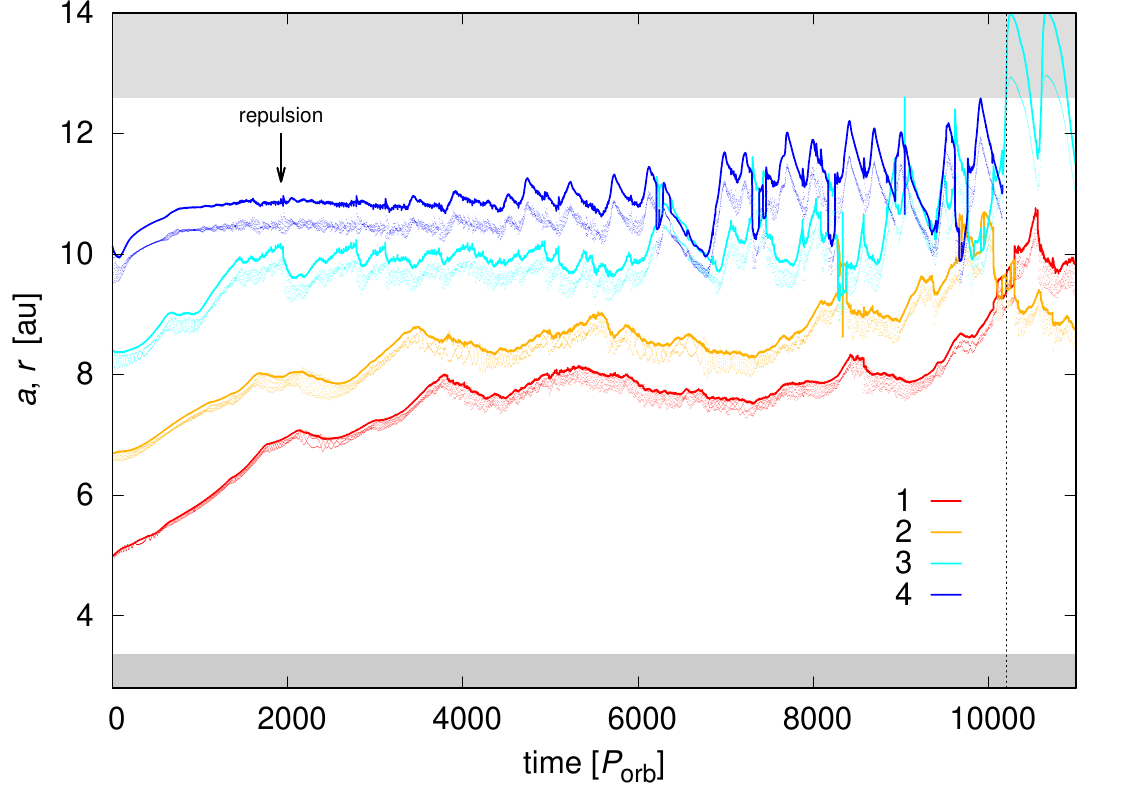}
\caption{The semimajor axis~$a$, and the heliocentric distance~$r$ vs time~$t$
for the simulation {\tt Sigma\char`_3times}.
The convergence radius is shifted further out to 11\,au.
Merging seems more difficult in this case
as there are many 'repulsions', especially between embryos 3~and 4.
The black arrow indicates the repulsion event we study in detail.
Note the semimajor axis is systematically offset from the radial distance, $a > r$,
because the orbits are substantially non-Keplerian.
After approximately $t \doteq 10000\,P_{\rm orb}$ some unwanted
interaction with the disk edge occurs and the evolution is no more reliable.
The masses reach $5$ to $7\,M_\oplus$ prior to this.
}
\label{nbody.orbits.at1}
\end{figure}

For the surface density 3~times larger than the nominal case,
as in the \verb|Sigma_3times| simulation,
the surface-density structures of the hot trail effect are not so pronounced,
owing to the larger thermal capacity of the gas (i.e. specific multiplied by~$\Sigma$).
The temperature excess reaches up to 10\,K, not 20\,K as before.
The 0-torque radius is located further out at approximately 11\,au.
The oscillations~$r(t)$ are relatively small, with the radial distance
systematically smaller than the semimajor axis, $r < a$,
because the interior disk mass $\int\Sigma(r) 2\pi r\d r \simeq 0.03\,M_\odot$ is no longer negligible.
This is also a notable case of non-Keplerian orbits, and a `false' osculating
eccentricity~$e > 0$, which shall not be used anymore.
Generally, the evolution seems slower, although the time span
in~Figure~\ref{nbody.orbits.at1} is almost three times longer
and the migration rate is comparable to the nominal case,
${\rm d}a/{\rm d}t \simeq 10^{-3}\,{\rm au}/P_{\rm orb}$.

Embryos do {\em not\/} interact so strongly, their orbits stay next to each other
for a prolonged period of time, likely because the hydrodynamical eccentricity damping is strong.
Sometimes, there is a reverse inward migration of the inner embryo~1 or~2,
inducing also larger oscillations of~$r(t)$.
Moreover, these excursions seem to be often out of phase,
and so they are not of a~resonant, but rather of a~hydrodynamical origin.

Interestingly, there are more than 10~attempts of embryo~3 or~4
to enter the coorbital region of each other. More specifically, there are in total
17~repulsions,
20~exchanges,
2~temporary coorbitals, and eventually
1~merger,
again soon after an encounter with another embryo.
The last part is already affected by interactions with the disk edge
and the damping zone, which kills the outer spiral arm.

\subsubsection{Detail: Repulsion}

\begin{figure*}
\centering
\tmp=3.7cm
\includegraphics[height=\tmp]{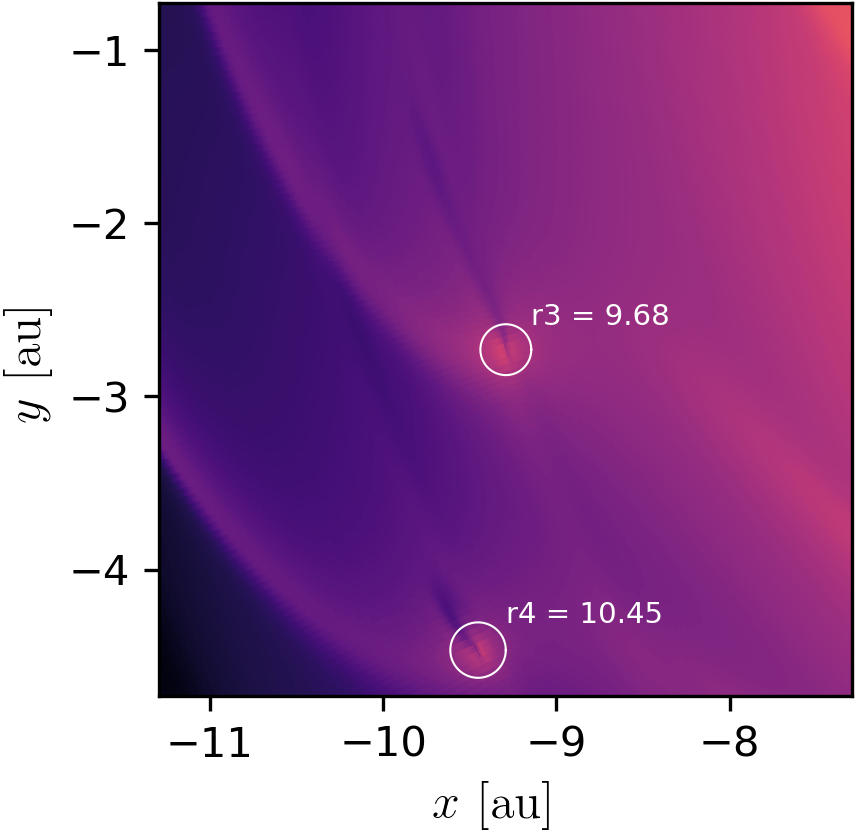}
\includegraphics[height=\tmp]{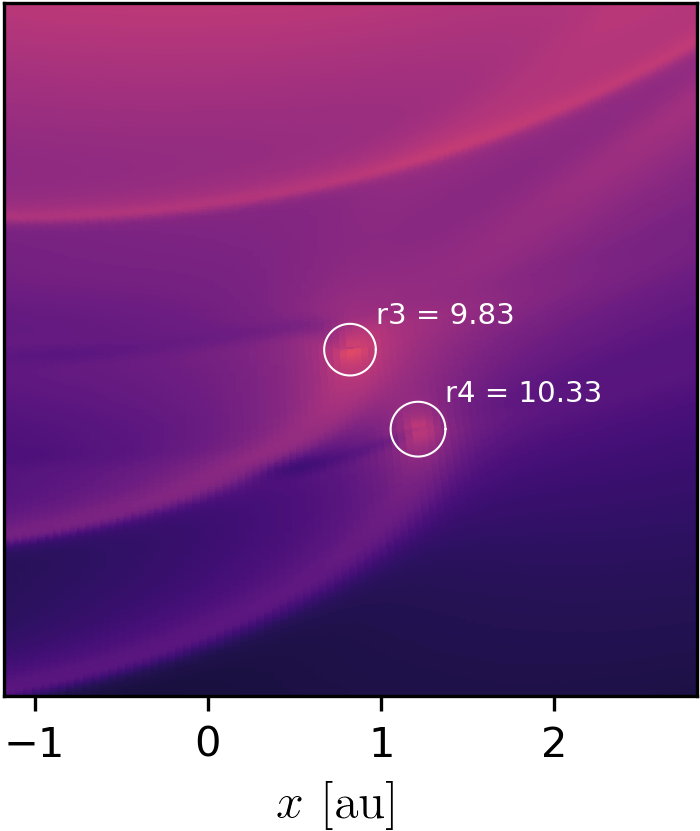}
\includegraphics[height=\tmp]{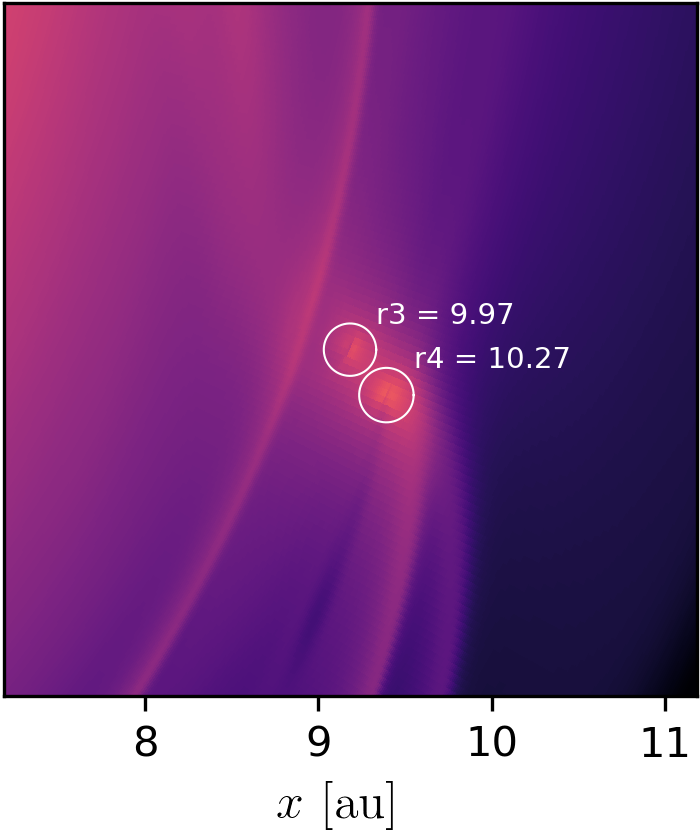}
\includegraphics[height=\tmp]{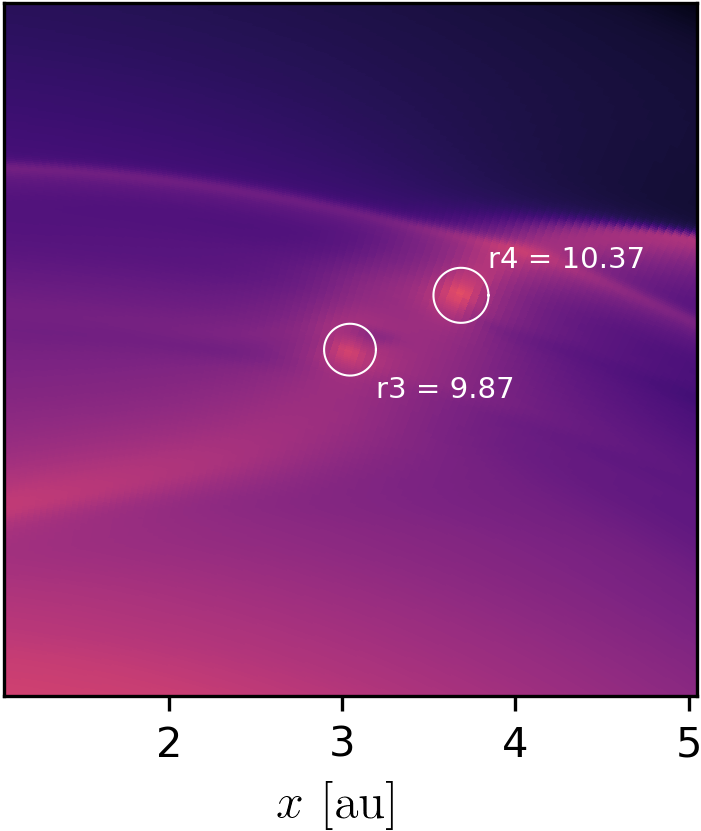}%
\includegraphics[height=\tmp]{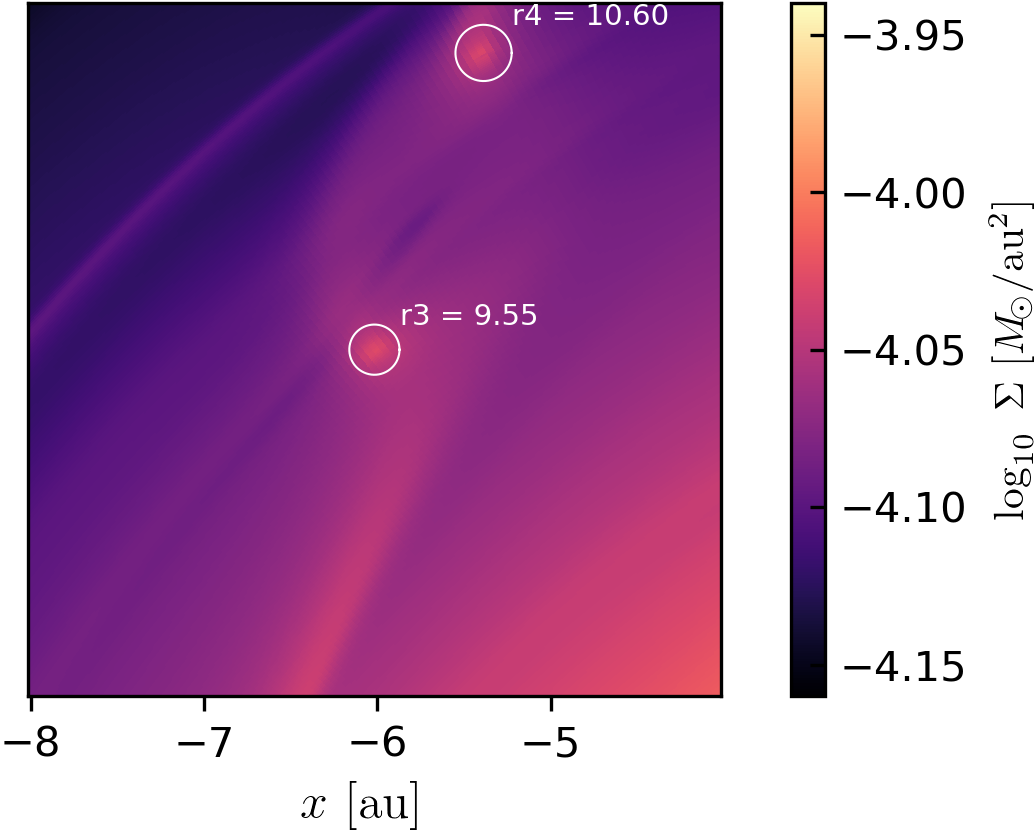}
\caption{The same as Fig.~\ref{CaseIII_nominal_Z1_MERGER1}
for a repulsion event in the {\tt Sigma\char`_3times} simulation
which occurred between $t = 1945$ to $1952\,P_{\rm orb}$.
The embryos approach each other in the apocentre and the pericentre
respectively, with spiral arms 'aligned'. There is an overdensity
between the Hill spheres during the closest approach,
and yet another spiral arm originating from the inner embryo~2 after the close approach.
}
\label{Sigma_3MMSN_Z1_REPULSE}
\end{figure*}

A detail of a~repulsion event is shown in Figure~\ref{Sigma_3MMSN_Z1_REPULSE},
namely the first one at $t \doteq 1945\,P_{\rm orb}$.
The embryos~3 and~4 approach each other in the apocentre and pericentre, respectively,
and their spiral arms are thus aligned. At the minimum distance
of just $2R_{\rm Hill}$, there is an overdensity between them.

The disk torque during this event is shown in Figure~\ref{tqwk},
in comparison with other types of events. It seems there are
alternating torques for the inner and outer embryos, which
contribute to the repulsion of the two. The next encounter
is consequently more distant.
An alternative Figure~\ref{tqwk_integral} expressed in terms
of the semimajor axis~$a(t)$ also shows the closest encounter itself
is driven by the interaction between the embryos, and the disk torque
acts as a perturbation. Nevertheless, the regular and repeated nature
of these repulsion events confirms that the torque actually determines
the encounter geometry.

\subsection{Low surface density}

\begin{figure}
\centering
\verb|Sigma_1over3|\break
\includegraphics[width=9cm]{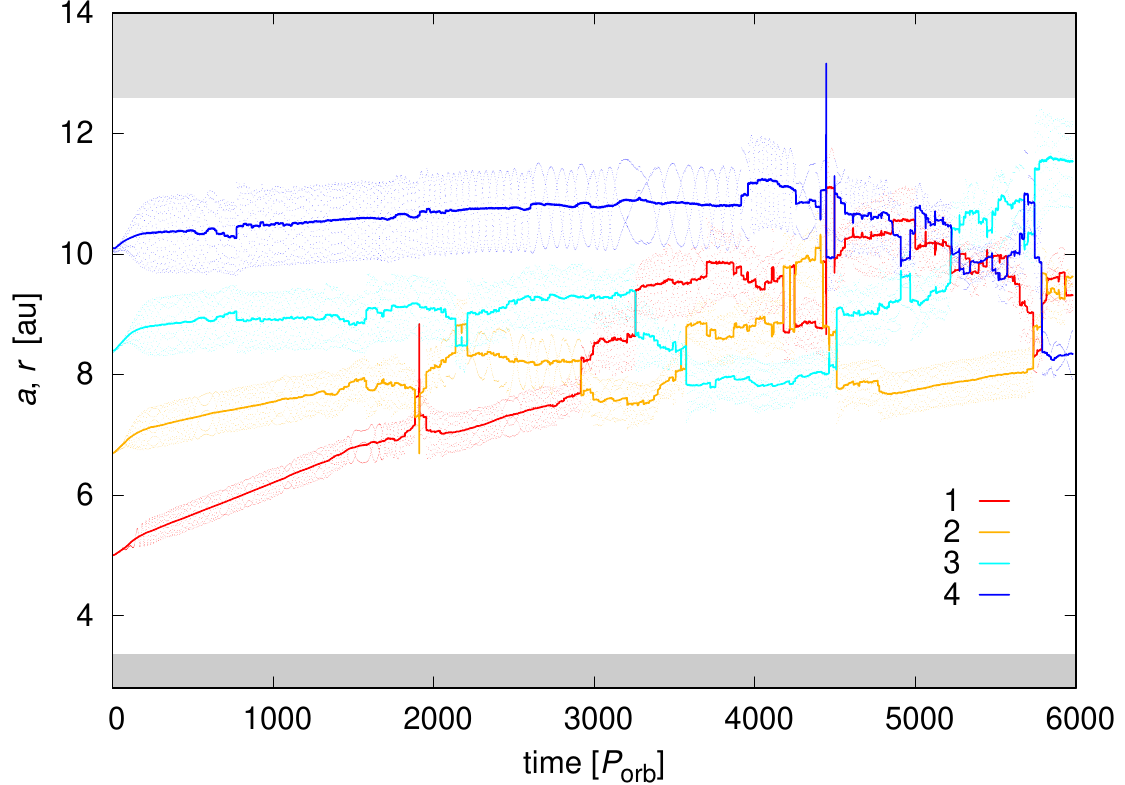}
\caption{The same as Fig.~\ref{nbody.orbits.at0}
for the simulation with $\Sigma_0$ three times smaller than the nominal one.
The oscillations~$r(t)$ induced by the hot-trail effect are larger
compared to the previous case, and also seem more stable.
There are numerous encounters in the evolution, but no mergers yet.
The final masses reach $5$ to $6\,M_\oplus$.
}
\label{nbody.orbits.at7}
\end{figure}

Generally, one would expect two limits exist:
for $\Sigma\to\infty$ the thermal capacity is so large, no hot-trial effect can develop;
for $\Sigma\to 0$ there is no hot-trail, because of no gas.
In the simulation \verb|Sigma_1over3|, we use the surface density
corresponding to 1/3 of the nominal (Figure~\ref{nbody.orbits.at7}).
It turns out the hot trail is even larger compared to the nominal case.
This is a result of smaller thermal capacity,
and also lower disk temperature which allows for a larger temperature excess.
Its development takes longer (more than $200\,P_{\rm orb}$),
and the trajectories seem to be more regular.

Overall the migration rate is comparable, although later
the motion often exhibits 'jumps', because the hydrodynamical damping is weaker.
There are:
28~exchanges,
9~repulsions (at least),
0~coorbitals, and
0~mergers in the course of the evolution.
Given this statistics, it is compatible to the nominal case
(after a normalisation to the same time span $4000\,P_{\rm orb}$);
coorbitals and mergers are a matter of small-number statistics, though.
One would need a sample of $10^2$ simulations to determine
which of the simulations would produce more of these relevant outcomes.

\subsection{Low pebble flux}

Yet another limit is clear: for $\dot M_{\rm p} \to 0$ the heating is zero
and the hot-trail effect too. The opposite case $\dot M_{\rm p} \to\infty$
is unclear (and unrealistic). For the simulation \verb|pebbleflux_2e-5|
we choose the pebble flux 10 times lower, corresponding to the embryo growth rate only $0.25\,M_\oplus$
per $4000\,P_{\rm orb}$. This may be actually more realistic
in later phases \citep{Lambrechts_Johansen_2014A&A...572A.107L};
the nominal case was on the high side, so we consider this simulation
to be potentially very important.

\begin{figure}
\centering
\verb|pebbleflux_2e-5|\break
\includegraphics[width=9cm]{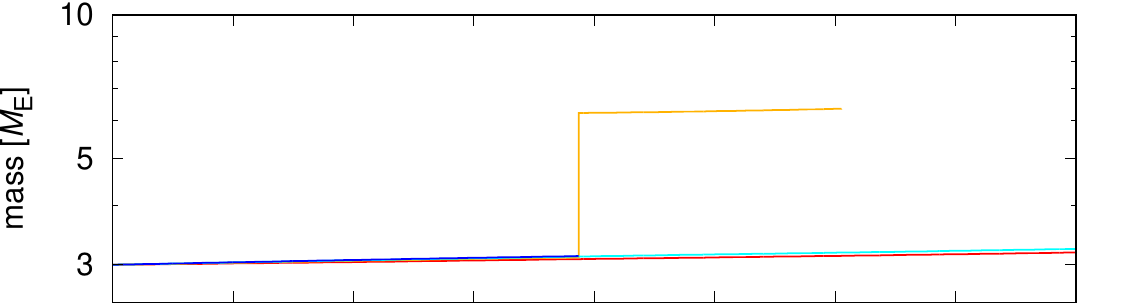}
\includegraphics[width=9cm]{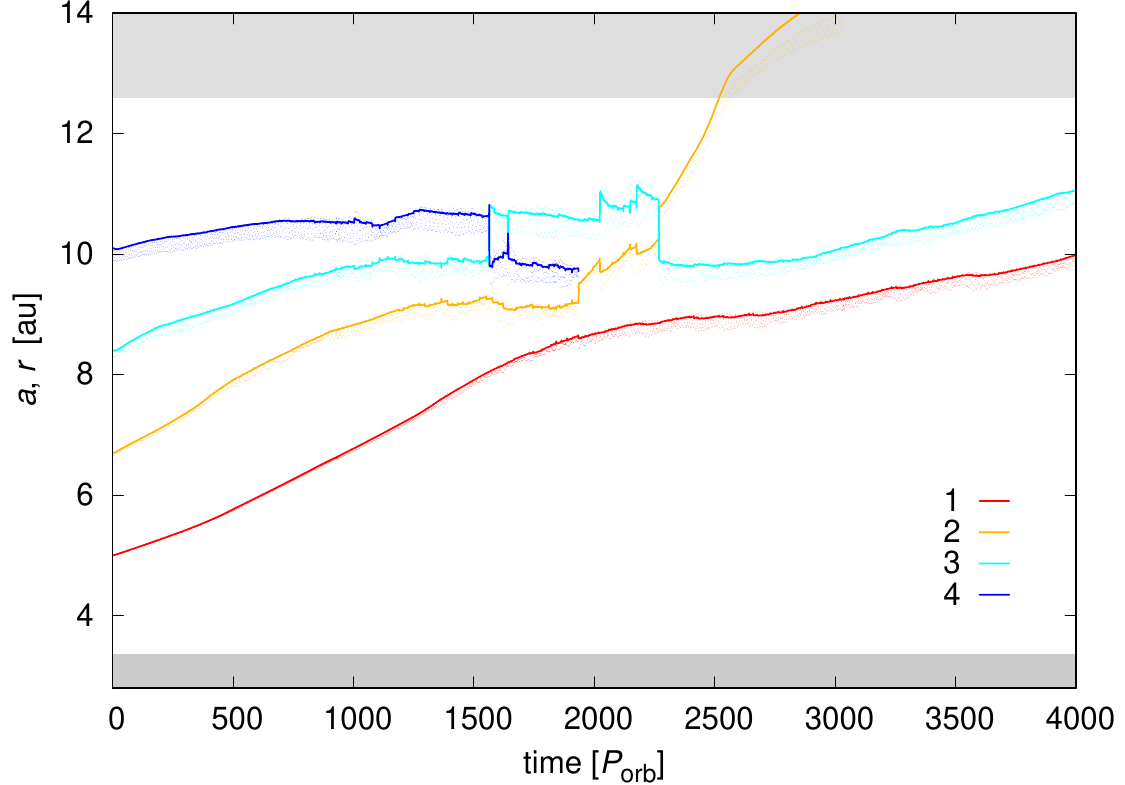}
\caption{The same as Fig.~\ref{nbody.orbits.at0}
for the pebble flux~$\dot M_{\rm p}$ ten times smaller than the nominal case.
The hot-trail effect ($e$'s) are then smaller than in the nominal case.
After a merger at $t \doteq 2250\,P_{\rm orb}$, the resulting $6M_\oplus$ embryo
quickly migrates outwards and is lost at the outer BC.}
\label{nbody.orbits.at4}
\end{figure}

As expected, hot-trail oscillations take longer to develop ($1000\,P_{\rm orb}$),
and they are about three times smaller afterwards (Figure~\ref{nbody.orbits.at4}).
The evolution is consequently more smooth, with no 'jumps', until the
embryos migrate to the convergence zone at about 11\,au.
This value seems somewhat surprising, because the heating
is substantially lower, and the 0-torque radius should be rather
smaller than the nominal one, $r < 9\,{\rm au}$. However,
an equilibrium (of Lindblad, corotation, and heating torques)
is perturbed here by the oscillations, and can be therefore
shifted elsewhere. For example, a small outward excursion takes
the embryo closer to the trailing spiral arm, without actually
crossing it, while a large excursion can reverse the Lindblad torque;
which is a bit counterintuitive.

The situation later dramatically changes, as there is
1~quick merger,
0~coorbitals,
2~exchanges,
2~big repulsions (plus many small),
because the torque acting on the $6\,M_\oplus$ merger is so strong,
it drives it outwards in a~runaway mode \citep{Pierens_Raymond_2016MNRAS.462.4130P}.
Obviously, this is the explanation for Planet~IX.
Taken more seriously, it turned out to be a rule for more massive embryos
(see Section~\ref{sec:totmass_20ME}).
This also leads to a clearing of the outer disk, beyond 11\,au.
The two remaining embryos~1 and~3 still migrate outwards,
their proper 0-torque radius being further out.
From the point of statistics, this simulation is not necessarily 
a representative one, and the early merger can be easily avoided
with a minor shift in ICs.

\subsection{Low viscosity}\label{sec:viscosity_1e-6}

\begin{figure}
\centering
\verb|viscosity_1e-6|\break
\includegraphics[width=8.5cm]{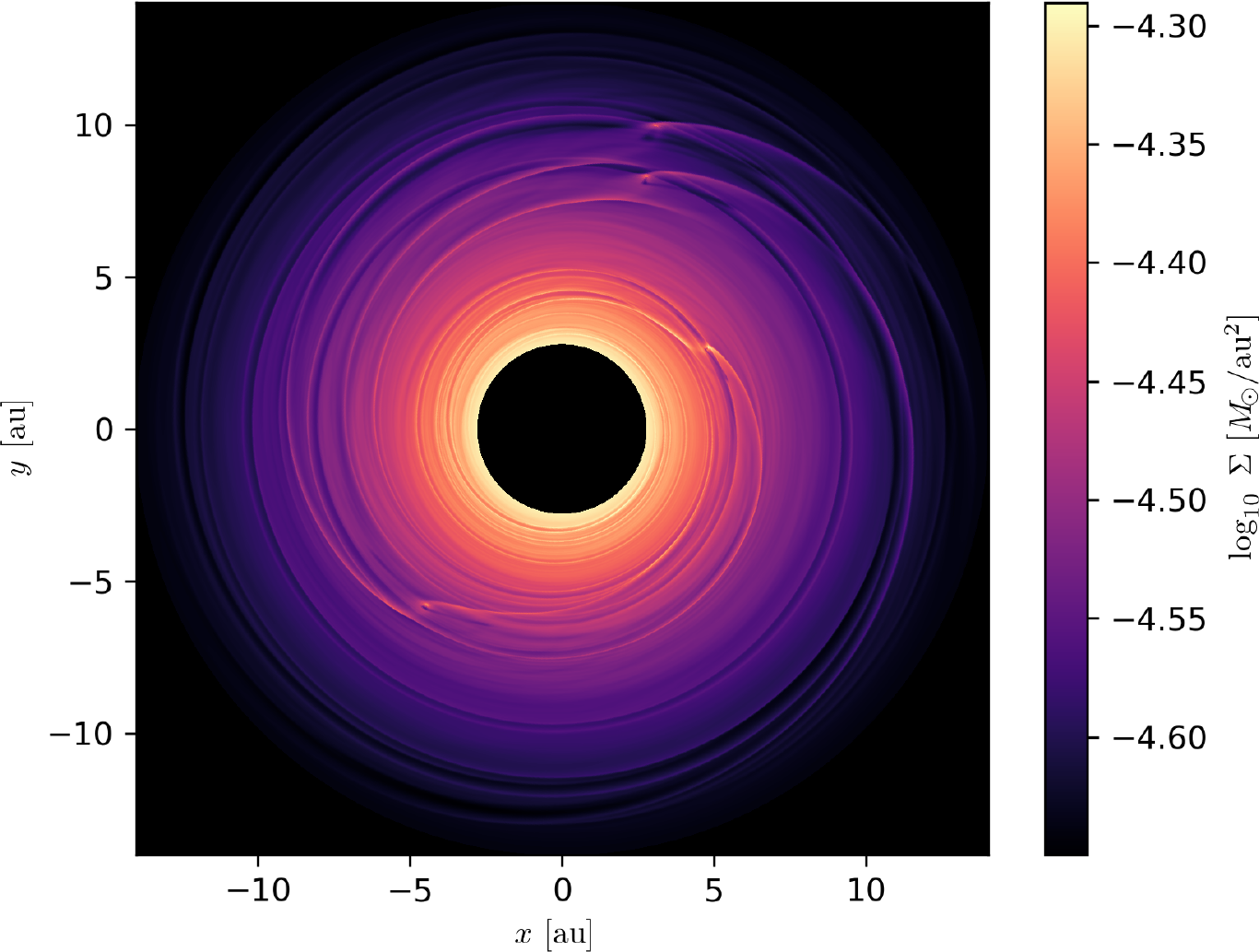}
\includegraphics[width=8.5cm]{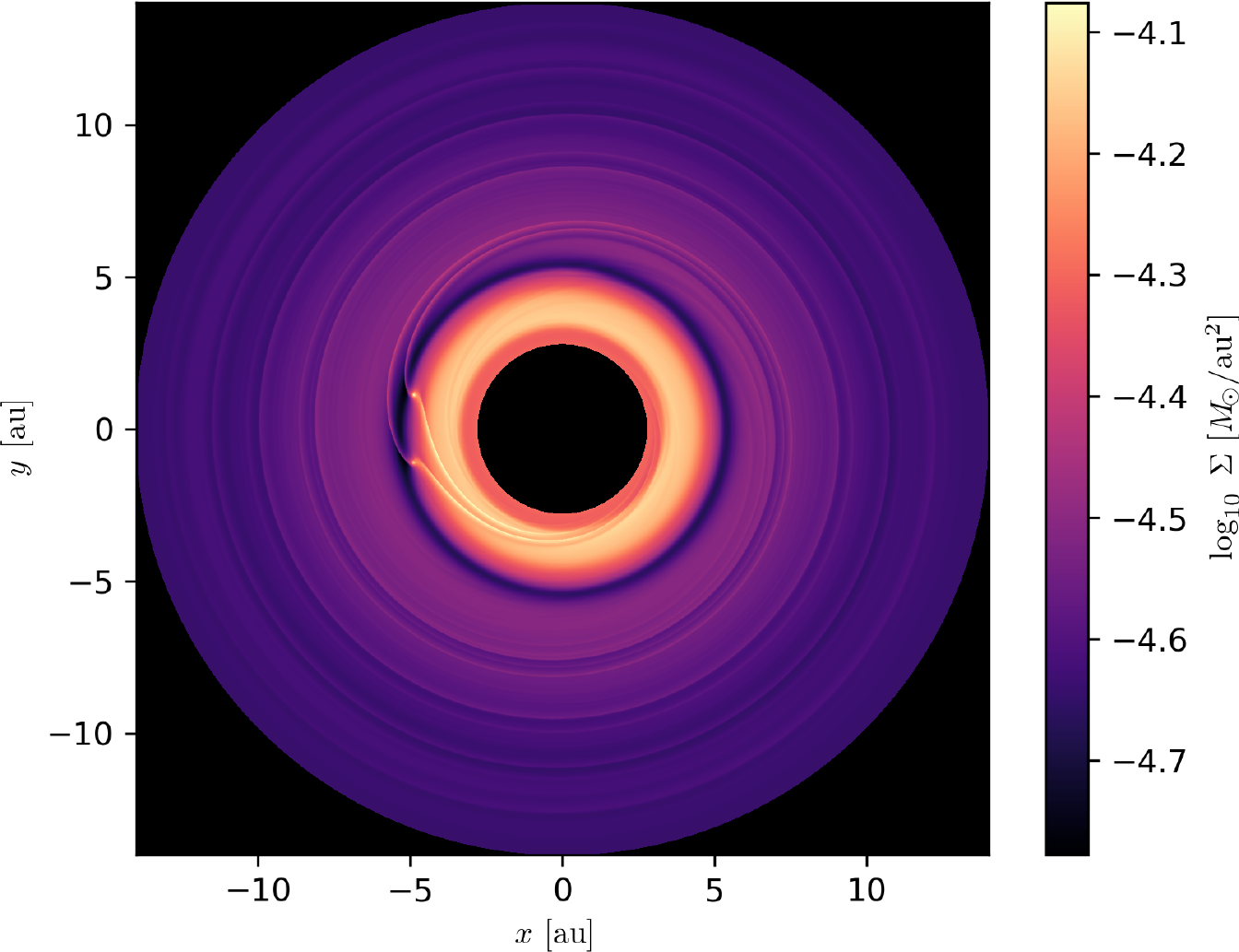}
\includegraphics[width=8.5cm]{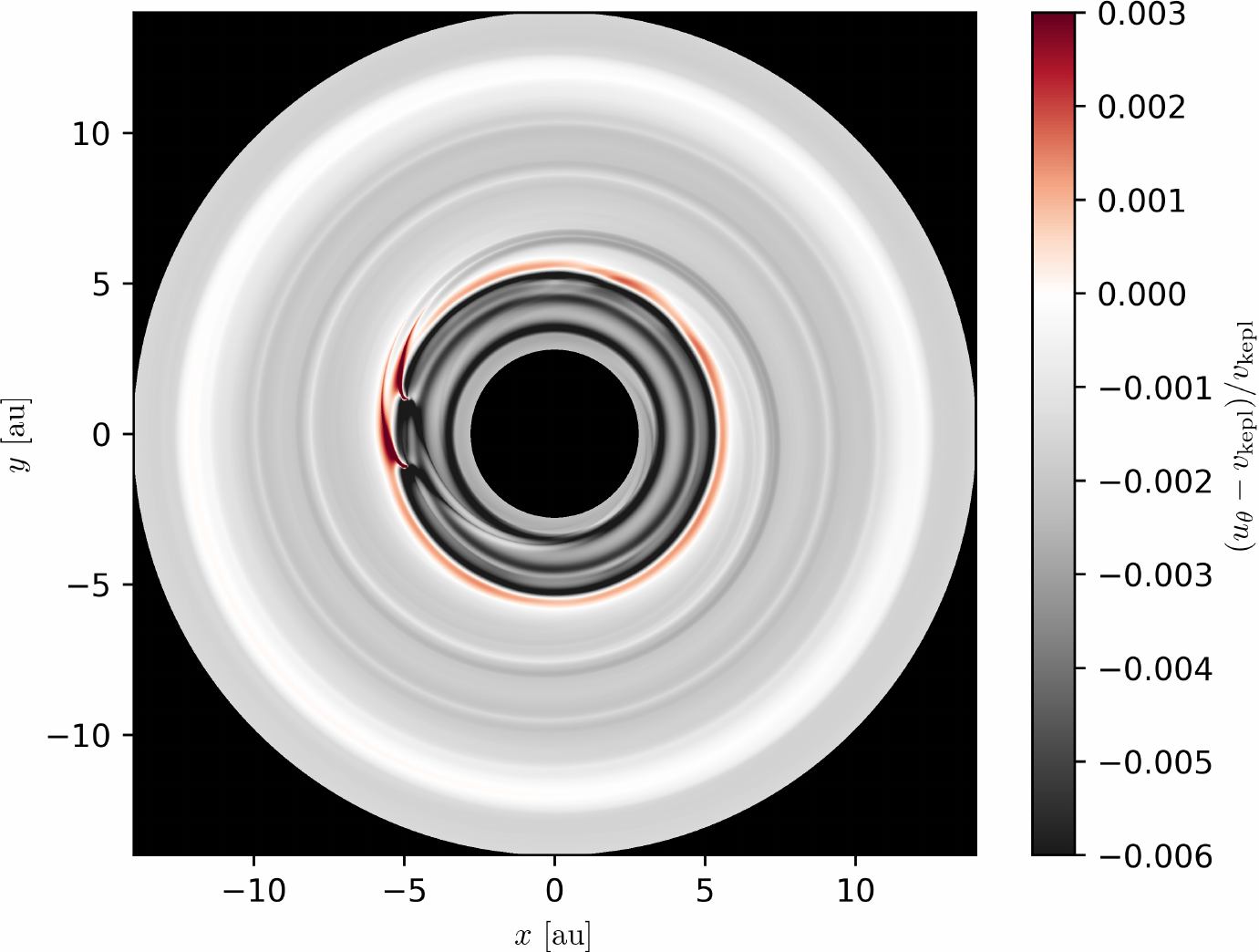}
\caption{The surface density~$\Sigma$ for the simulation
with ten times smaller viscosity~$\nu$;
the situation short after the beginning at $t = 100\,P_{\rm orb}$ (top),
and the final state at $t = 8000\,P_{\rm orb}$ (middle).
The structures are initially more pronounced compared to the nominal case,
because perturbations of the low-viscosity gas spread more slowly.
The outcome is a massive coorbital which migrates towards the inner boundary.
The density contrast between~$\Sigma$ interior and exterior to the coorbital
is about~2; the disk is optically thick in both cases.
The corresponding azimuthal velocity~$(u_\theta - v_{\rm kepl})/v_{\rm kepl}$
of pebbles (bottom), compared to the Keplerian velocity.
Just outside the coorbital pair $u_\theta$ is positive
(super-Keplerian) and a pebble isolation develops.
}
\label{gasdens4_6}
\end{figure}

As discussed in the Introduction,
the disk -- or its dead zone with a negligible ionisation and no MRI turbulence --
could have been also (almost) inviscid. To this point, we performed a simulation
with 10~times lower viscosity (codenamed \verb|viscosity_1e-6|).
The structures in the gas disk shown in Figure~\ref{gasdens4_6}
are more pronounced;
the viscous spreading is diminished, thus any density perturbation
arising from protoplanets persists longer.

\begin{figure}
\centering
\verb|viscosity_1e-6|\break
\includegraphics[width=9cm]{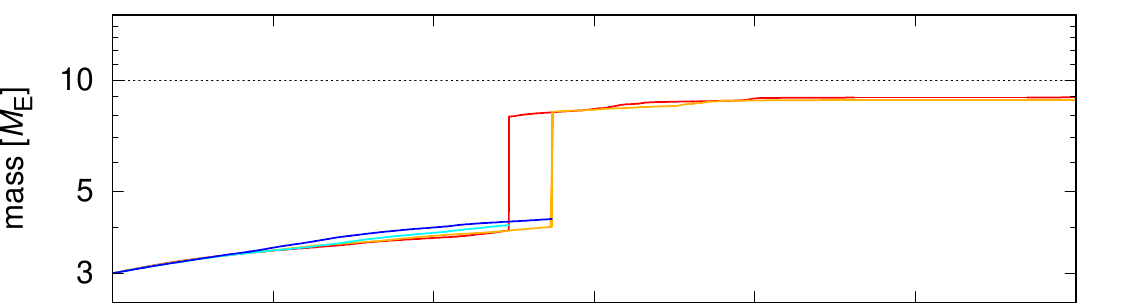}
\includegraphics[width=9cm]{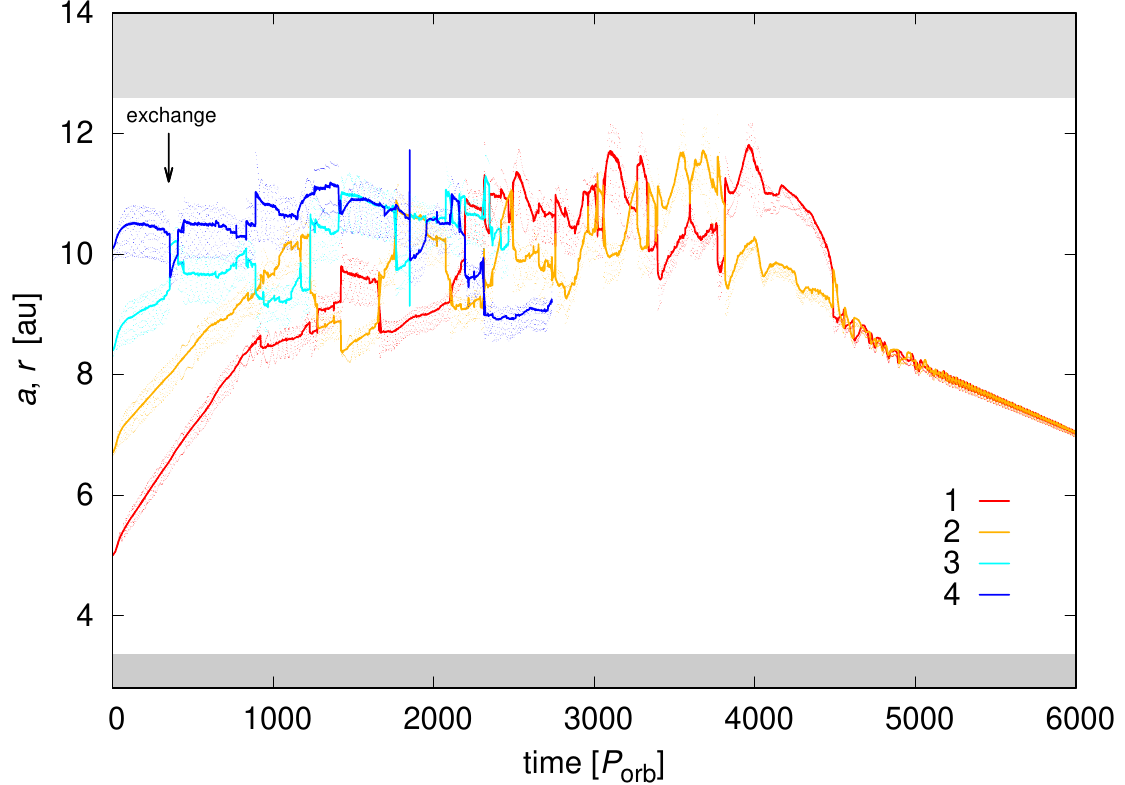}
\caption{The same as Fig.~\ref{nbody.orbits.at0}
for the simulation with ten times smaller viscosity~$\nu$.
The oscillations~$r(t)$ are initially similar to the nominal case,
but there is a faster migration towards the convergence radius.
After many 'trials', there are two mergers ($8\,M_\oplus$),
which later form the coorbital pair.
The black arrow indicates the repulsion event we study in detail.}
\label{nbody.orbits.at6}
\end{figure}

The evolution itself (Figure~\ref{nbody.orbits.at6}) indicates
the same hot-trail oscillations, but a faster migration towards
the 0-torque radius \citep{Paardekooper_etal_2011MNRAS.410..293P}.
Embryos pushed by larger torques then interact
more violently and this results in:
18~exchanges,
19~repulsions,
2~mergers -- created by 3-body interactions -- and
5~coorbitals, in total.
Out of these, 4~are only temporary, which experience another 3-body interactions,
during which exchange their members (around $t \doteq 2000\,P_{\rm orb}$).
The last one, formed by two~$8M_\oplus$ embryos is stable,
and is further stabilized.

This stabilisation is facilitated by a pebble isolation
which develops beyond the coorbital pair
(Fig.~\ref{gasdens4_6}, bottom;
seen as a stalled accretion $\dot M_{\rm em} \doteq 0$ in Fig.~\ref{nbody.orbits.at6}, top).
It also implicitly drives the migration inwards; it switches off accretion heating
by isolating the coorbital from further solid material.
The pair is massive enough to create a surface density contrast of about~2
between the interior and exterior part of the gas disk,
or 2.5 if we take the middle of the gap.
Nevertheless, both are still optically thick; it is not yet a gap opening.
Unless there is a substantial filtering of dust \citep{Rosotti_etal_2016MNRAS.459.2790R},
or an extensive shadow hiding the outer part,
or very long wavelengths ($\lambda\gg a_{\rm dust}$) are used,
it would be difficult to observe.

\subsubsection{Detail: Exchange}

\begin{figure*}
\centering
\includegraphics[height=\tmp]{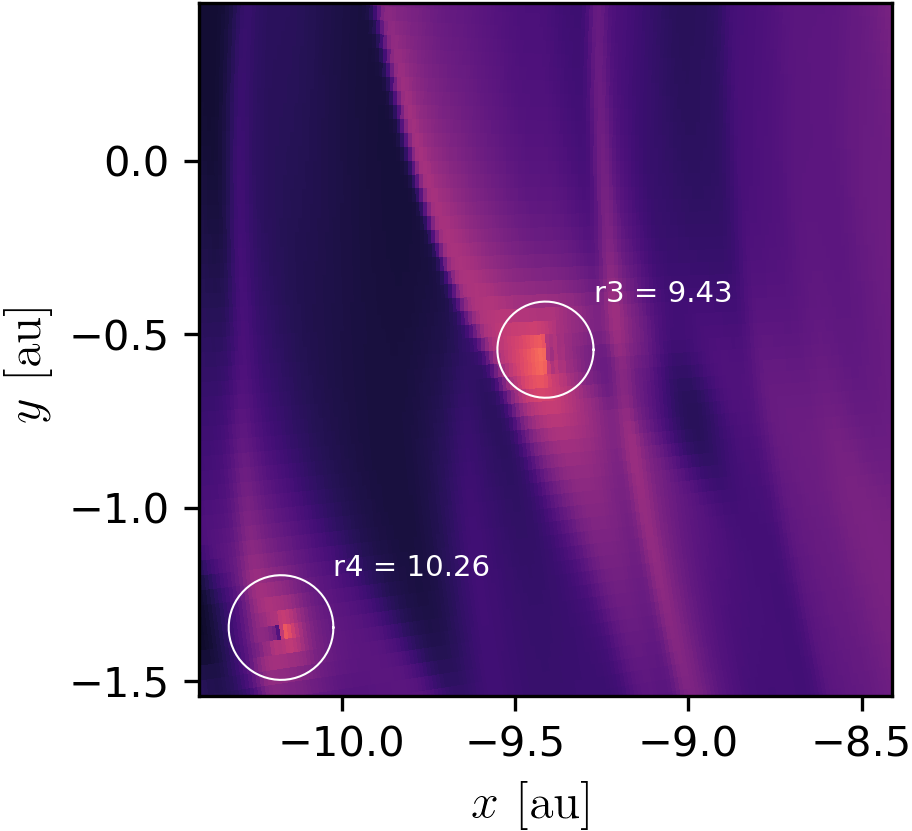}%
\includegraphics[height=\tmp]{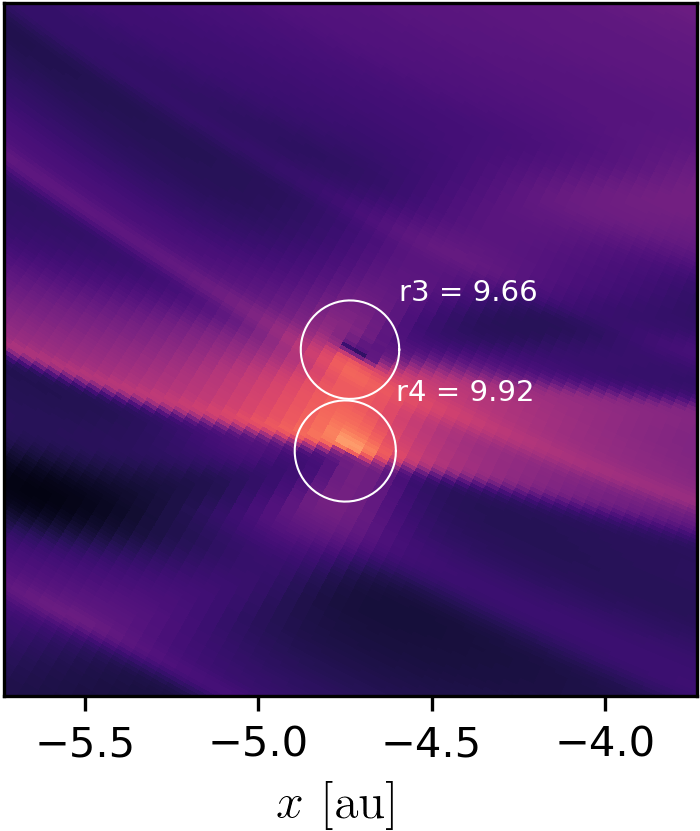}
\includegraphics[height=\tmp]{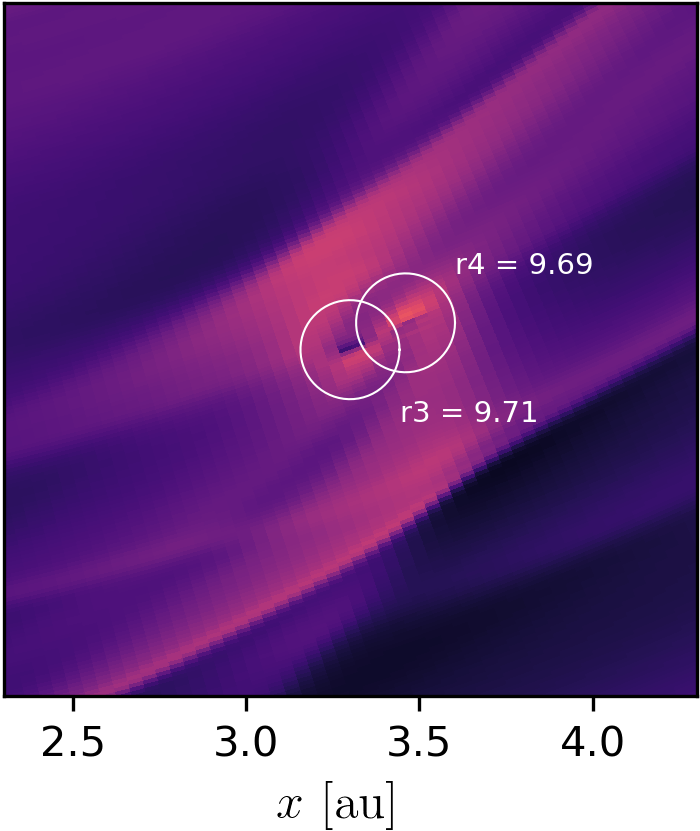}
\includegraphics[height=\tmp]{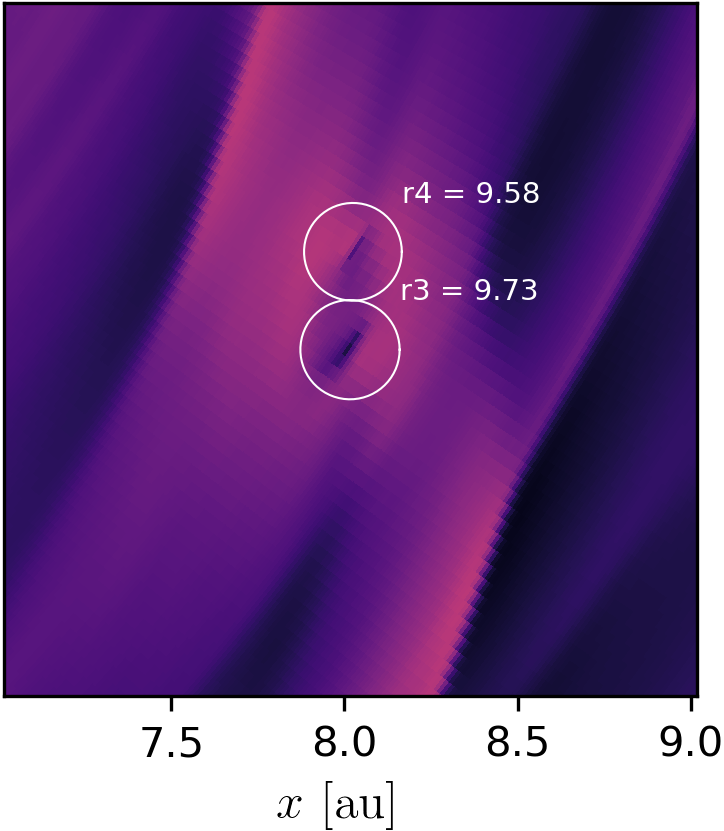}%
\includegraphics[height=\tmp]{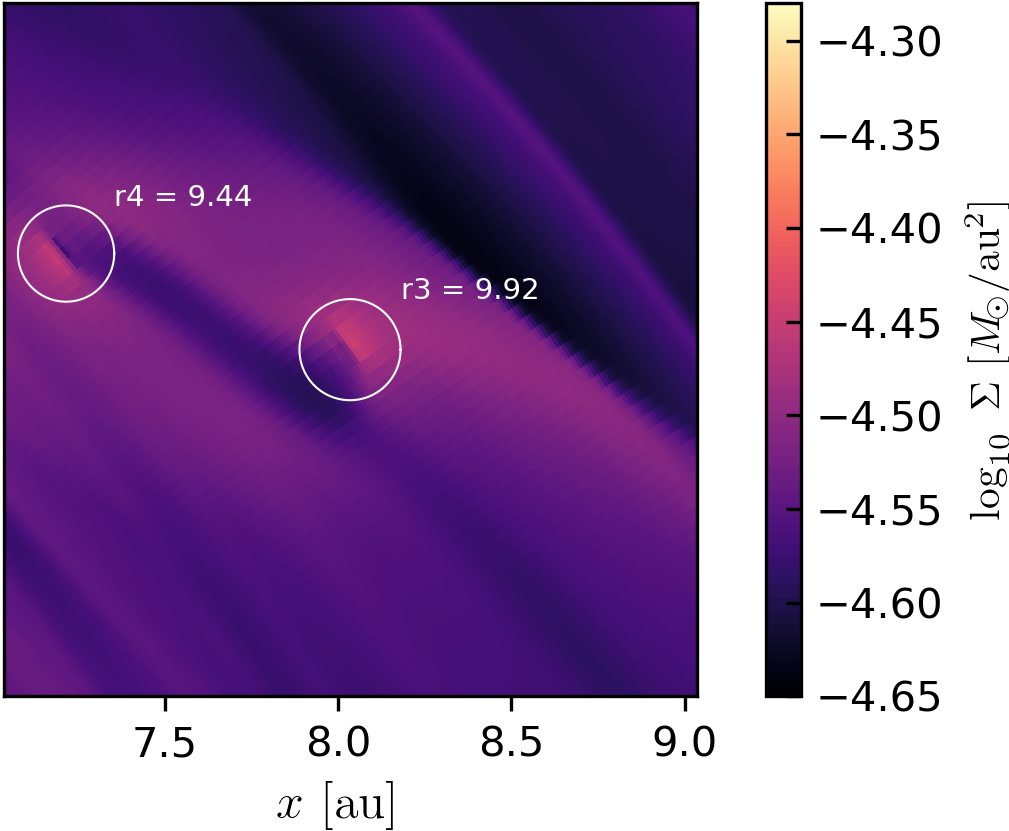}
\caption{The same as Fig.~\ref{CaseIII_nominal_Z1_MERGER1} for an exchange event
in the {\tt viscosity\char`_1e-6} simulation. The embryos~3 and~4 approach
in such a way, the inner flies behind the outer, and a shared underdense
region is formed which connects the embryos during their retreat.
}
\label{viscosity_1e-6_Z2_EXCHANGE}
\end{figure*}

A~detail of the first exchange is shown in Figure~\ref{viscosity_1e-6_Z2_EXCHANGE}.
Embryo~3 at the apocentre encounters embryo~4 at the pericentre.
An overdense region is formed between them as the outer and inner
spiral arms overlap. During the closest approach at a distance of $1.2\,R_{\rm H}$,
the gas distribution is uneven. As the embryos become more distant,
an extended underdense region is formed between. Disk torques contribute
to the exchange, being mostly negative on the outer embryo and positive
on the inner one (Fig.~\ref{tqwk}). Apart from these major perturbations,
there are many more minor density waves created by other (inner) embryos.

\subsection{Gas accretion}

As a test, we performed also a simulation with the gas accretion
and the corresponding heating switched on, but the efficiency parameter
was very small, $f_{\rm acc} = 10^{-6}$, as we started with $3\,M_\oplus$ bodies.
Specifically, the gas accretion rate is given by the integral
over the exponential density profile \citep{Kley_1999MNRAS.303..696K}:
\begin{eqnarray}
\left({\pd\Sigma\over\pd t}\right)_{\rm acc} &=& \sum_i {1\over3}f_{\rm acc} \int\!\!\!\int {\cal H}(|\vec r_{\perp i}-\vec r_{\rm cell}|-0.75 R_{{\rm H}i})\,\times \nonumber\\
 && \times\, {\Sigma\over\sqrt{2\pi}H} \int_{z_i-\Delta z}^{z_i+\Delta z}\! \exp\left[-\left({z\over\sqrt{2}H}\right)^2\right] \d z\,\d\theta\d r\,,
\end{eqnarray}
plus the same term with numerical factors ${2\over3}$ and 0.45,
where ${\cal H}$ denotes the Heaviside step function,
$\vec r_\perp$ the planet position in $(x,y)$ plane,
$\Delta z = \sqrt{(0.75R_{\rm H})^2-|\vec r_\perp\!-\!\vec r_{\rm cell}|^2}$, and
$H$ the vertical scale height.

\begin{figure}
\centering
\verb|gasaccretion_1e-6|
\vskip.1cm
\includegraphics[width=9cm]{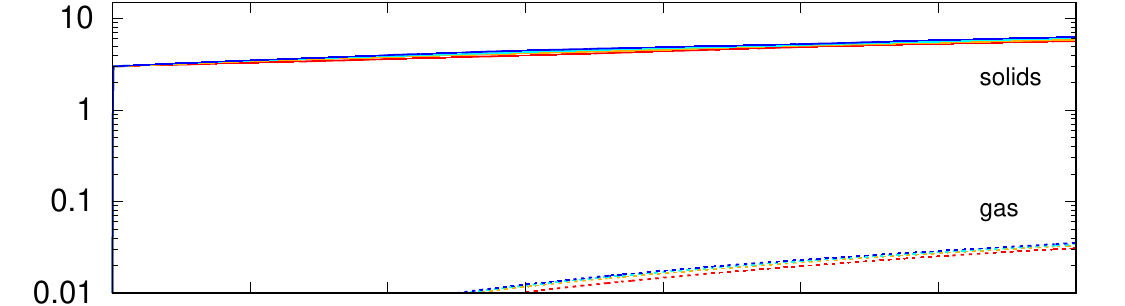}
\includegraphics[width=9cm]{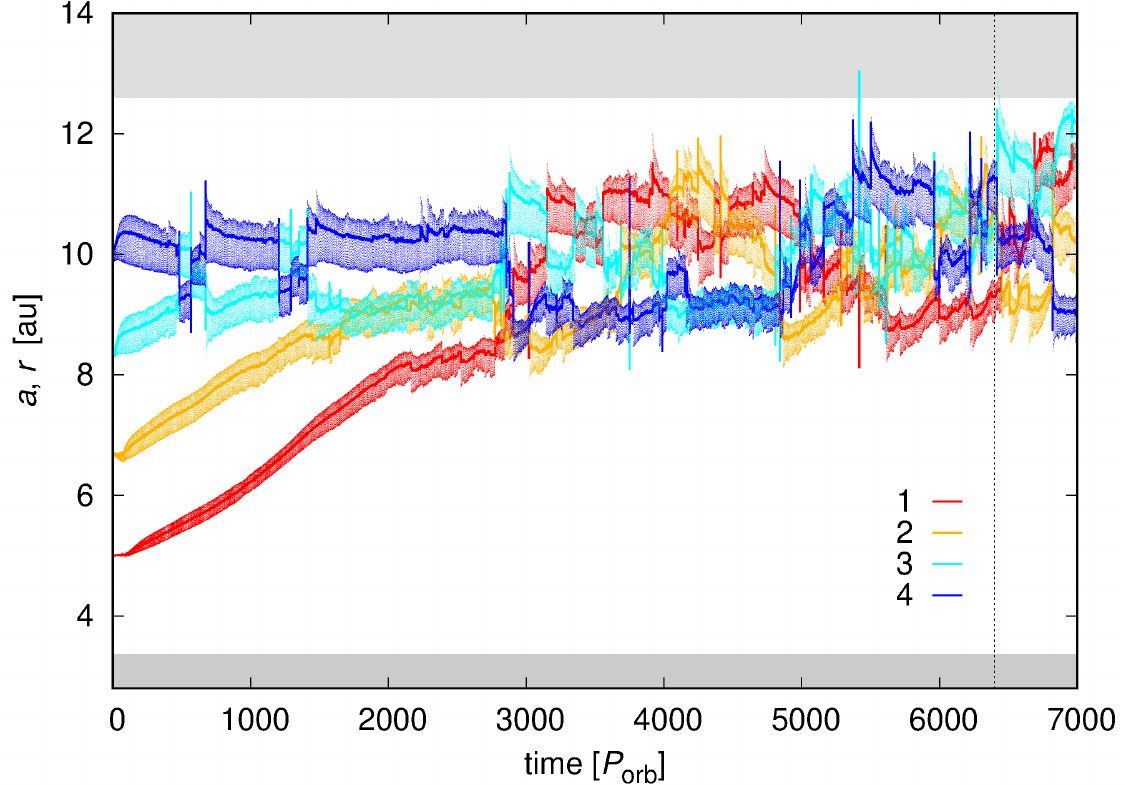}
\caption{The semimajor axis~$a$ (line), and the heliocentric distance~$r$ (dots) vs time~$t$ (bottom),
and the embryo mass~$M_{\rm em}$ vs~$t$ (top) for the simulation
including the gas accretion, with the efficiency factor~$f_{\rm acc} = 10^{-6}$.
We distinguish the solid (solid) and gaseous (dashed) component of~$M_{\rm em}$.
Note we used a~finer sampling of the orbital elements.
Both the migration rates and the eccentricities seem the same
as in the nominal case, because the gas accretion contributes
only very little to the mass and heating.
However, no merging occurred in the course of this simulation.
After $t \doteq 6400\,P_{\rm orb}$ there is an interaction
with the disk edge and the evolution is no more reliable.}
\label{fargo_gasacc_nbody.orbits.at5}
\end{figure}

The total amount of gas thus reaches at most $0.035\,M_\odot$.
We may regard this simulation more as another realisation of the nominal one.
Indeed, according to Figure~\ref{fargo_gasacc_nbody.orbits.at5},
the hot-trail oscillations are practically the same, and the migration
rates too. The difference stems from the chaotic nature of the system.
A comparison of the first $4000\,P_{\rm orb}$ shows a similar frequency
of the common events, albeit there are no coorbitals or mergers yet,
which are simply not frequent enough to occur in every simulation.
We did not continue the simulation because of the interactions with
the outer damping zone.

\subsection{Four $5M_\oplus$ embryos}\label{sec:totmass_20ME}

\begin{figure}
\centering
\verb|totmass_20ME|\break
\includegraphics[width=9cm]{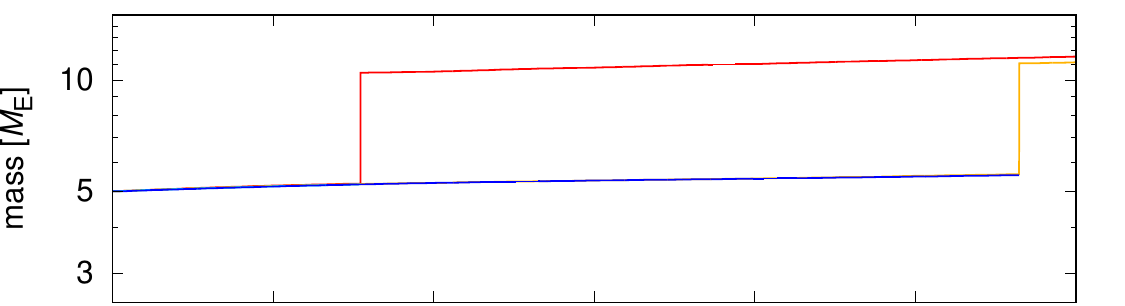}
\includegraphics[width=9cm]{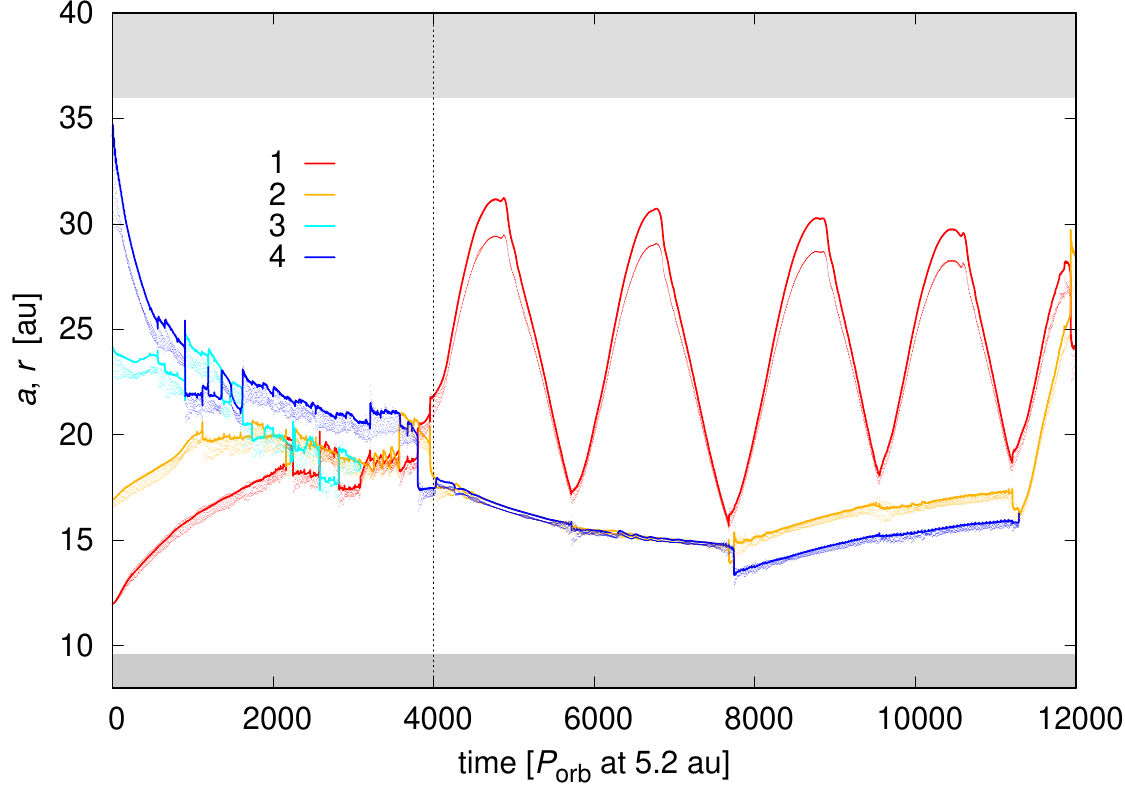}
\caption{The same as Fig.~\ref{nbody.orbits.at0} for the simulation
with $5M_\oplus$ embryos. We model an outer disk here,
because the convergence radius turned out to be far (21\,au).
After numerous exchanges, temporary coorbitals, and one merger,
a relatively stable coorbital is formed at $t \doteq 4000\,P_{\rm orb}$.
Embryo~1 is then pushed outwards by a dynamical torque created
by an underdense tadpole-like region \citep{Pierens_2015MNRAS.454.2003P}.
At $t \doteq 4900\,P_{\rm orb}$ the tadpole region is refilled 
by a material originating from the inner spiral arm,
and embryo~1 is pushed inwards, until it encounters the coorbital.
This behaviour is systematically repeated,
and results in a~disruption of the coorbital, and a~merger.}
\label{nbody.orbits.at5}
\end{figure}

With initial masses $5\,M_\oplus$ all embryos quickly drifted outwards.
The 0-torque is clearly much more distant, and our setup requires an
outer disk spanning from~8 to~40\,au (i.e. \verb|totmass_20ME|).
In other words, this may be relevant for the formation of ice giants.
The spacing of embryos is 16 mutual Hill radii, similarly as before.
We also deliberately decreased the pebble flux to $\dot M_{\rm p} = 10^{-5}\,M_\oplus\,{\rm yr}^{-1}$,
because larger embryo masses are usually attained later.
A convergence to the new 0-torque radius at about 20\,au is relatively
fast, especially when longer orbital periods are taken into account
(Figure~\ref{nbody.orbits.at5}).

The evolution is similarly complex as in the inner disk, with
20~exchanges,
5~repulsions (not counting the small ones),
2~mergers, and
2~coorbitals,
with the last one (at $t \doteq 4000\,P_{\rm orb}$) stabilised;
this time it is not due to the pebble isolation,
but rather pebble filtering due to embryo~1,
presently the outer massive merger.

A very interesting behaviour then starts: embryo~1 is pushed outwards
by a dynamical torque created by an underdense tadpole-like region \citep{Pierens_2015MNRAS.454.2003P}.
Later at $t \doteq 4900\,P_{\rm orb}$, when embryo~1 reaches
its new convergence radius~$r_{\rm c}$, it slightly overshoots,
is pushed backwards and the tadpole region is refilled 
by a material originating from the inner spiral arm.
With no underdensity anymore, embryo~1 is pushed inwards,
until it encounters the coorbital.
This behaviour is systematically repeated four times,
with a disruption of the coorbital in the meantime,
and a second merger after all.

\subsection{Eight $1.5M_\oplus$ embryos}

\begin{figure}
\centering
\verb|embryos_1.5ME_8|\break
\includegraphics[width=9cm]{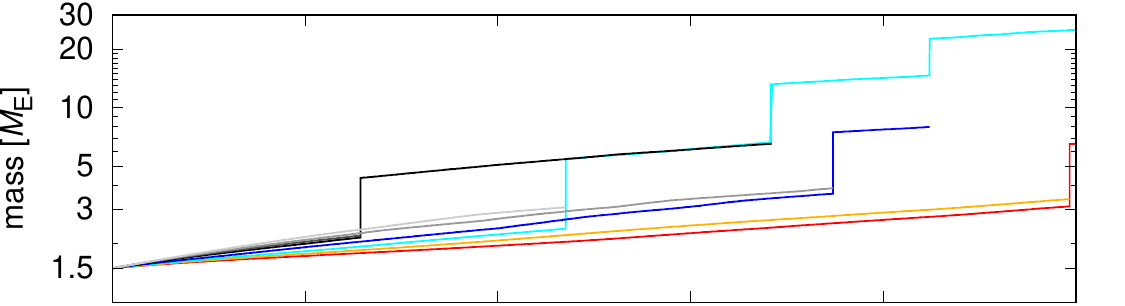}
\includegraphics[width=9cm]{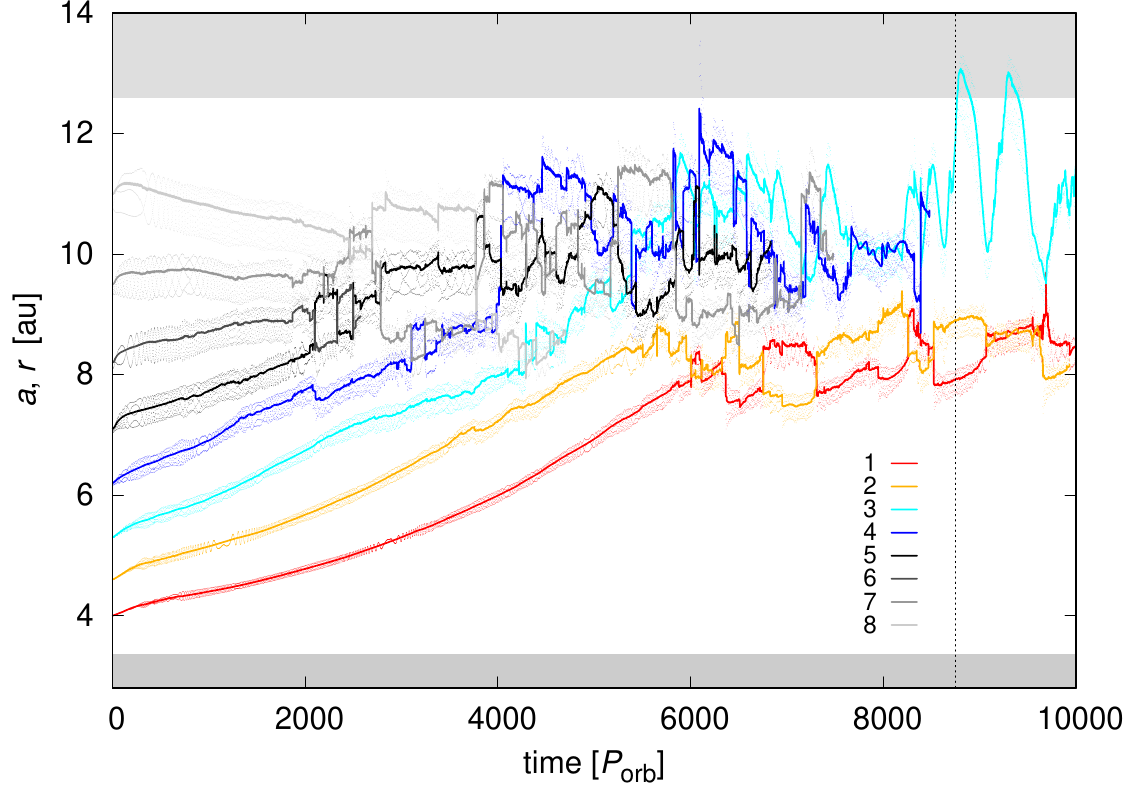}
\caption{The same as Fig.~\ref{nbody.orbits.at0} for the simulation
with eight $1.5M_\oplus$ embryos. The evolution seems qualitatively
similar to the nominal case, with 6 merger events amidst.
Most if not all mergers occur when 3-body interactions take place.
After $t \doteq 8750\,P_{\rm orb}$ an interaction of the outer embryo
with the disk edge occurs and the evolution is no more reliable.}
\label{nbody.orbits.at3}
\end{figure}

For eight embryos with $1.5\,M_\oplus$ each, we can see a clear convergence
to the 0-torque radius at approximately 9\,au (as in the nominal case). The oscillations due to
the hot-trail are apparently small for the innermost embryos,
but they become larger as they migrate outwards. At the given radial
distance, the amplitudes are the same, of course (Figure~\ref{nbody.orbits.at3}).

The migration is slower compared to the nominal case,
nevertheless, the initial spacing is more compact (10~mutual Hill radii),
and interactions start at about the same time.
They are of all kinds, both two-body and three-body,
but we will not count them explicitly,
as the number of objects is twice larger.
Generally, there are more opportunities to merge,
ending up in 5~mergers, the most massive having up to $25\,M_\oplus$.
From this standpoint, the simulation can be termed as successful,
creating a giant-planet core with more than a critical mass.
Our current model is not reliable for this large mass though,
as we used no Hill cut,
zero gas accretion efficiency factor~$f_{\rm acc}$,
and there are problems to describe the pebble isolation in 2D \citep{Bitsch_etal_2018A&A...612A..30B}.
A~sixth merger occurred after a series of unwanted interactions
with the outer damping zone, and this makes further evolution also unreliable.

In this simulation (and also in \verb|viscosity_1e-6|), the embryos
gain non-negligible temporary inclinations of the order of $10^{-4}\,{\rm rad}$,
despite of \cite{Tanaka_Ward_2004ApJ...602..388T} damping.
The vertical distances are then orders of magnitude larger
than protoplanet diameters, which decreases collisional probabilities.
On the other hand, a gravitational focussing
(with $v_{\rm esc} \doteq 7\,{\rm km}\,{\rm s}^{-1}$) --
which is accounted in our model -- helps to counteract it.

\subsection{Many low-mass embryos}

Finally, we simulated a system composed of 120 embryos with $0.1\,M_\oplus$ each
(i.e. \verb|embryos_0.1ME_120|); the total mass remains the same as the nominal.
We chose a tight spacing of 2~mutual Hill radii:
\begin{equation}
R_{\rm HH} = {1\over 2}(a+a')\left({q+q'\over 3}\right)^{1/3},
\end{equation}
with $a$ the semimajor axis and $q$ the planet-to-star mass ratio,
to fit all bodies in a disk spanning from 2.8 to 16 au.
The initial state was already shown in Figure~\ref{gasdens4_2}.
A convergence test for a single embryo
is presented in Appendix~\ref{sec:convergence}.
Our current resolution is still low, 3~cells per Hill sphere.
The test shows that $\d a/\d t$ is then overestimated by a factor of~3,
and $r(t)$ oscillations (or eccentricity~$e$) have the same amplitude.
We consider this as an acceptable approximation which may somewhat
enhance the efficiency of merging (due to larger $\d a/\d t$).
Having correct eccentricities seems more important to prevent
spurious resonant captures. It is a very slow computation anyway,
with 120 disk$\,\leftrightarrow\,$planet interactions;
to this point it was run on the NASA Pleiades supercomputer,
with $10^3$ reserved CPU cores, and a combined MPI/OpenMP parallelisation.
To better resolve the Hill sphere, one would have to increase
the number of cells in {\em both\/} radial and tangential directions,
fulfill the Courant condition in these smaller cells, and compute
all the interactions, so it scales almost as $N^3N_{\rm em}$.

\begin{figure}
\centering
\includegraphics[width=9cm]{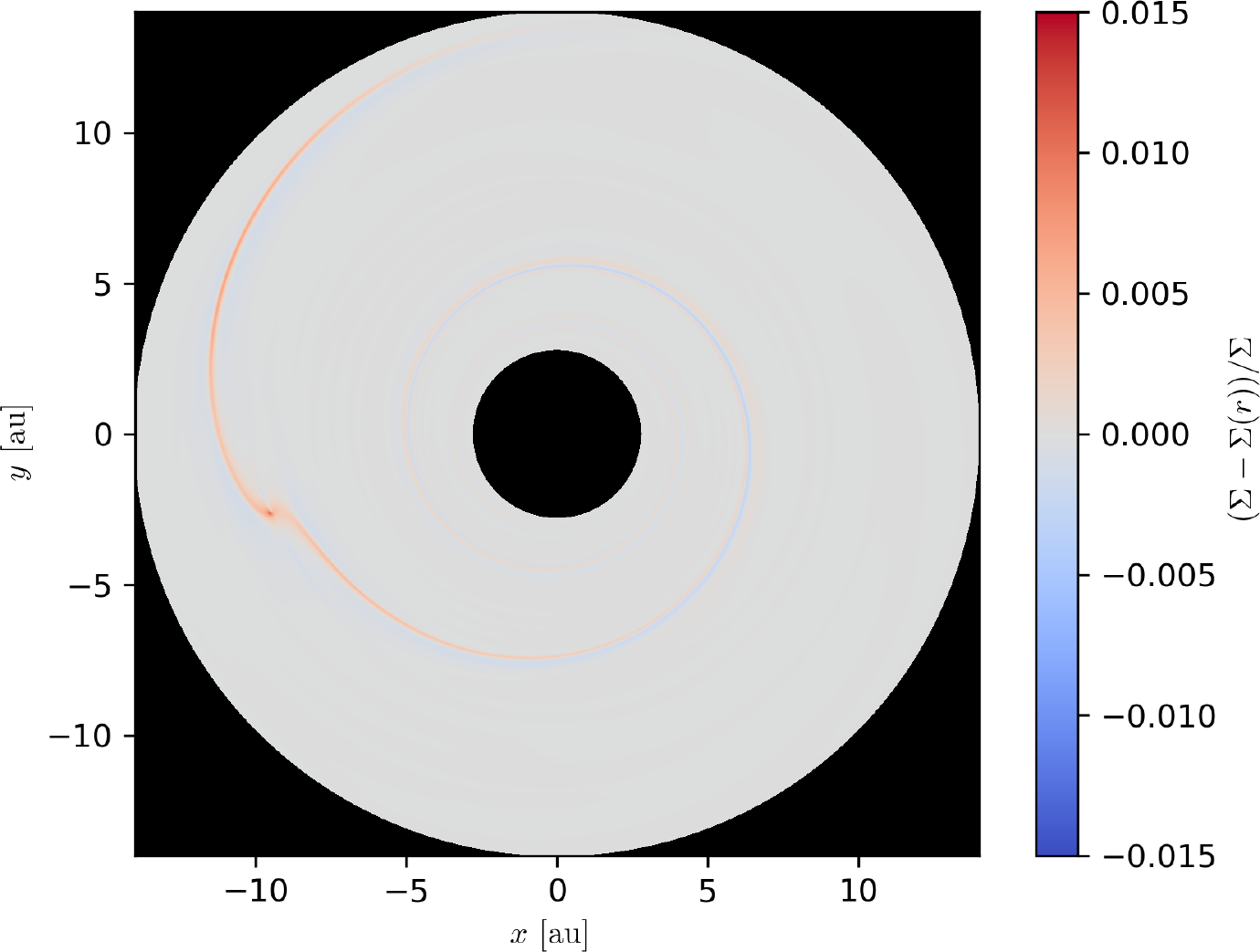}
\includegraphics[width=9cm]{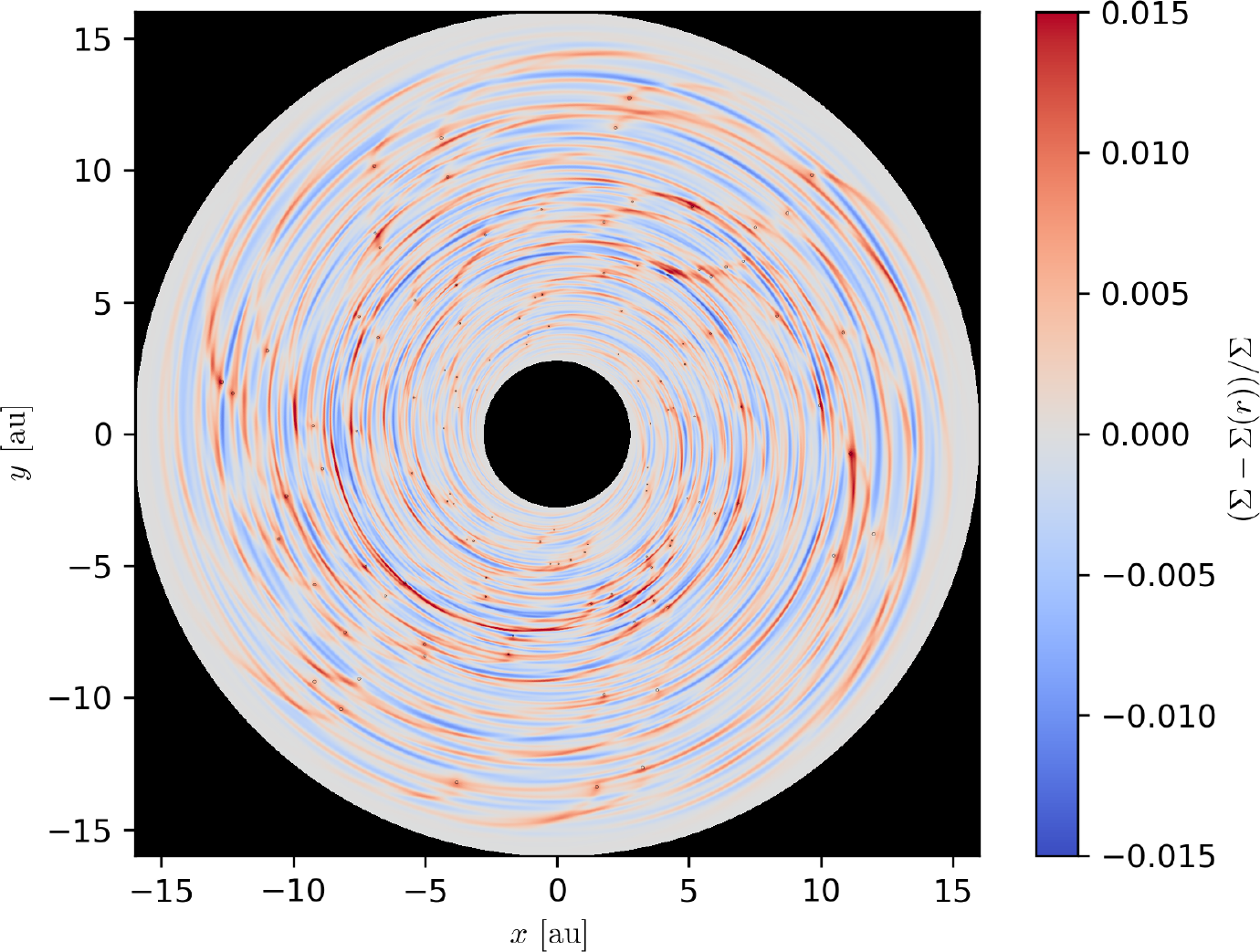}
\caption{The surface density~$\Sigma(r,\theta)$ of the gas disk
with the azimuthally averaged profile $\Sigma(r)$ subtracted
to clearly see the respective spiral arms, the corotation region,
and other perturbations the surroundings of the Hill sphere.
The system was evolved for 100 orbital periods $P_{\rm orb}$ (at 5.2\,au)
so that the hot trail effect can develop and increase the eccentricity.
The situation is very different for a single $0.1\,M_\oplus$ embryo (top),
with a very regular spiral,
and for 120 embryos with the same masses (bottom),
with spiral arms overlapping each other,
and creating an irregular overall pattern.
The situation corresponds to Fig.~\ref{gasdens4_2},
but it is much easier to see the perturbations when $\Sigma-\Sigma(r)$ quantity is used.
The resolution $3072\times 4096$ was used for the former
short-term simulation, and $2048\times 3072$ for the latter.
The Hill spheres are shown as small black circles.
}
\label{migra4_gasdens_SPIRAL}
\end{figure}

The most important result is visible already at the beginning --
there is no smooth Type-I migration, because there are no regular patterns
(see Figure~\ref{migra4_gasdens_SPIRAL})!
The weak spiral arms overlap, and affect neighbouring corotation regions too.
These {\em stochastic torques\/} are very different from
the classical, regular, single-planet torques.
Although Type-I migration in its simplest form may be considered linear,
it is no longer true, because our system of equations
(Eqs.~(\ref{eq:dSigma_dt}) to (\ref{eq:ddot_r}))
includes several non-linear terms ($T^3$, $T^4$, $Q_{\rm visc}$).
Mutual gravitational interactions of the embryos also partly
contribute to the stochastic nature of the system, especially given
the initial spacing ($2\,R_{\rm HH}$).

\begin{figure}
\centering
\verb|embryos_0.1ME_120|\break
\includegraphics[width=9cm]{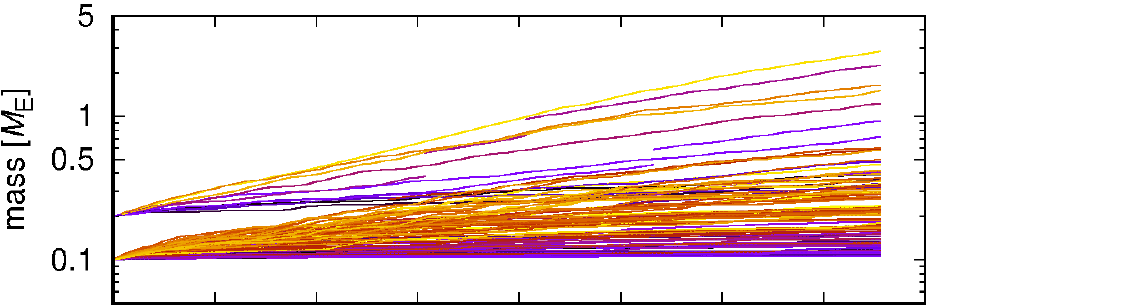}
\includegraphics[width=9cm]{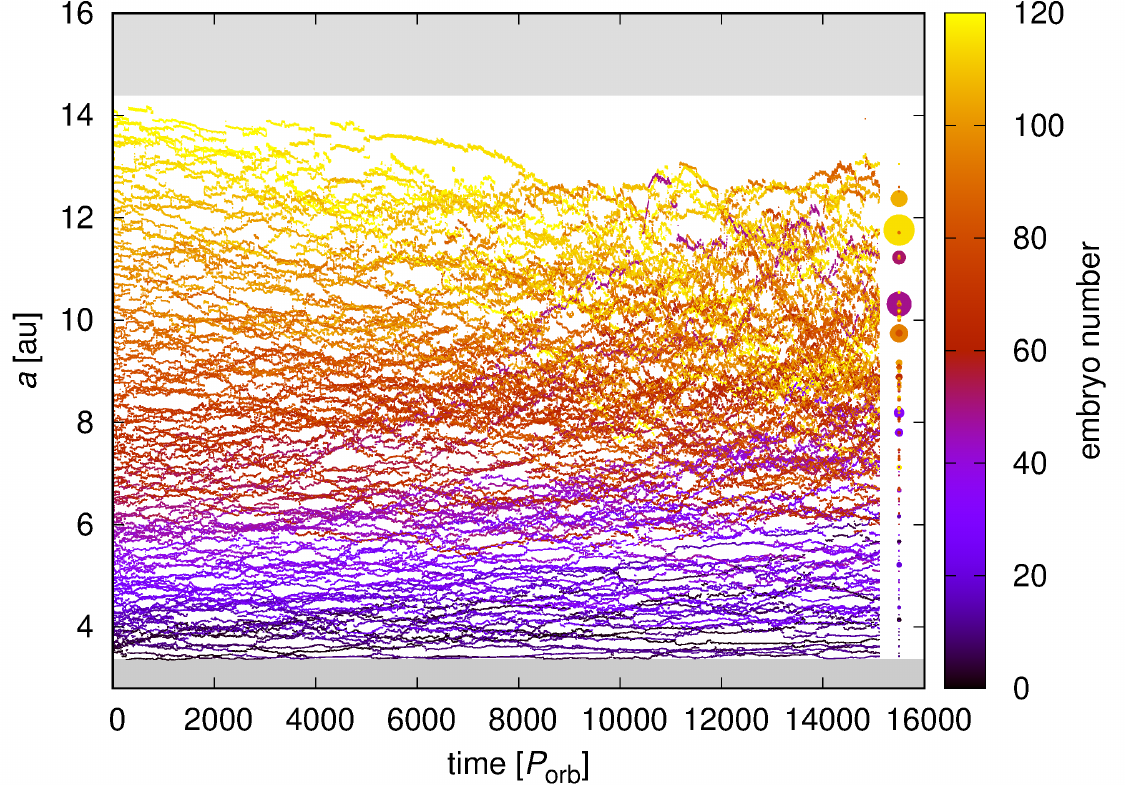}
\caption{The semimajor axis~$a$ vs time~$t$ (bottom),
and the embryo mass~$M_{\rm em}$ vs~$t$ (top) for the simulation
with 120 low-mass $0.1M_\oplus$ embryos. Colours correspond
to the embryo number to distinguish the individual orbits.
The final state is depicted as a series of filled circles,
with sizes proportional to the masses.
The evolution is never regular but rather chaotic,
partly due to direct N-body gravitational interactions
among the embryos, but more importantly due to overlapping
spiral arms (see also Fig.~\ref{migra4_gasdens_SPIRAL}).
Initially, there are only several mergers which create
a handful of $0.2M_\oplus$ embryos. These grow preferentially
by the pebble accretion; there are a~few additional mergers.}
\label{nbody.orbits.at2}
\end{figure}

The evolution of the whole system is surprisingly slow (Figure~\ref{nbody.orbits.at2}).
The oscillations induced by the hot-trail effect ($e\doteq 0.02$) serve
as an initial 'kick', which leads to numerous close encounters.
At the same time, inclinations~$i$ are excited too,
being virialised with the eccentricities~$e$,
The average values are different though, $e\simeq 0.02$, $i\simeq 0.01\,{\rm rad}$,
because \cite{Tanaka_Ward_2004ApJ...602..388T} damping acts on inclinations.
This is a new situation; in all previous simulations
inclinations remained low (less than $10^{-3}$ even during close encounters).
Given the increase of~$e$,
we expected there will be many mergers early in the simulation,
but this is suppressed by the increase of~$i$.
There are only 11 of them which occur during the first
few~$100\,P_{\rm orb}$, creating a group of $0.2M_E$ embryos,
located both in the inner and outer parts of the disk.
It might be an artifact of our initial conditions,
but it does not 'hurt' us in any way, because having some
range of masses initially seems even more realistic.

\begin{figure}
\centering
\includegraphics[width=8.5cm]{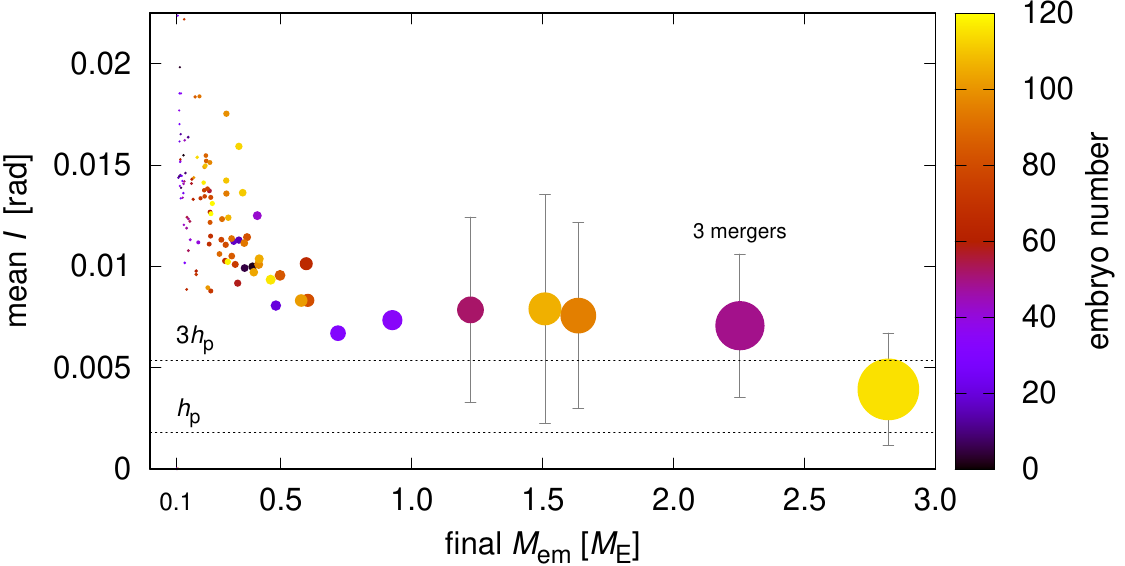}
\caption{The mean inclination~$\bar I$ (radians) versus the final embryo
mass~$M_{\rm em}$. The symbol sizes correspond also to $M_{\rm em}$
and their colours to the embryo number, or its initial position.
The horizontal dotted lines indicate multiples of the pebble scale
height, $h_{\rm p} = H_{\rm p}/r$. The smaller vertical distance
of the massive embryos from the pebble disk contributes to a runaway growth.
All embryos are already in the Hill regime,
and their effective accretion radius ranges from
$R_{\rm eff}/r = 0.003\hbox{ to }0.012$.}
\label{embryos_0.1ME_120_meani}
\end{figure}

Further growth is facilitated mostly by the pebble accretion.
Only a handful of additional mergers occur (see Fig.~\ref{nbody.orbits.at2}, top).
We identified five processes which contribute to a runaway growth:
  (i)~the initial mergers ($0.2\,M_\oplus$), other embryos have a~mass `handicap';
 (ii)~in the Hill regime of pebble accretion the cross section is proportional to the embryo mass,
      $\pi R_{\rm H}^2 \propto M_{\rm em}^{\rm 2/3}$,
      and the relative velocity too,
      $v_{\rm H} \propto R_{\rm H} \propto M_{\rm em}^{1/3}$,
      which results in an exponential $M_{\rm em}(t)$ evolution;
(iii)~late mergers of inner embryos occur, which subsequently drift outwards;
 (iv)~pebble filtering by outer (already massive) embryos;
  (v)~a separation between low- and high-mass embryos, with the former
      having larger {\em mean\/} inclinations than their pebble feeding zone
      (Figure~\ref{embryos_0.1ME_120_meani}).
The peak masses are almost $3\,M_\oplus$ for the group of early mergers.
while only $0.6\,M_\oplus$ for the group of original embryos.

\begin{figure}
\centering
\verb|embryos_0.1ME_120|\break
\includegraphics[width=8.5cm]{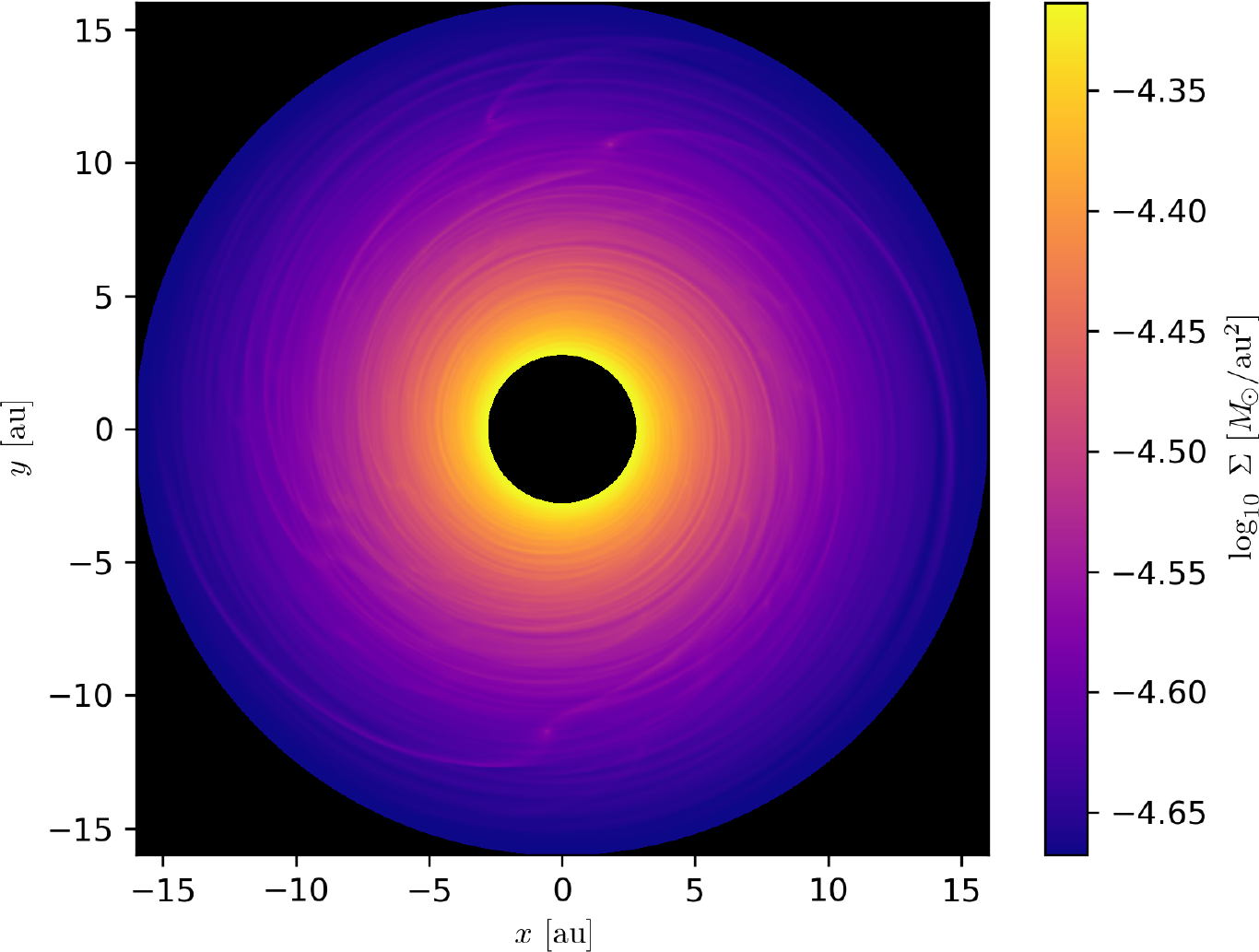}
\includegraphics[width=8.5cm]{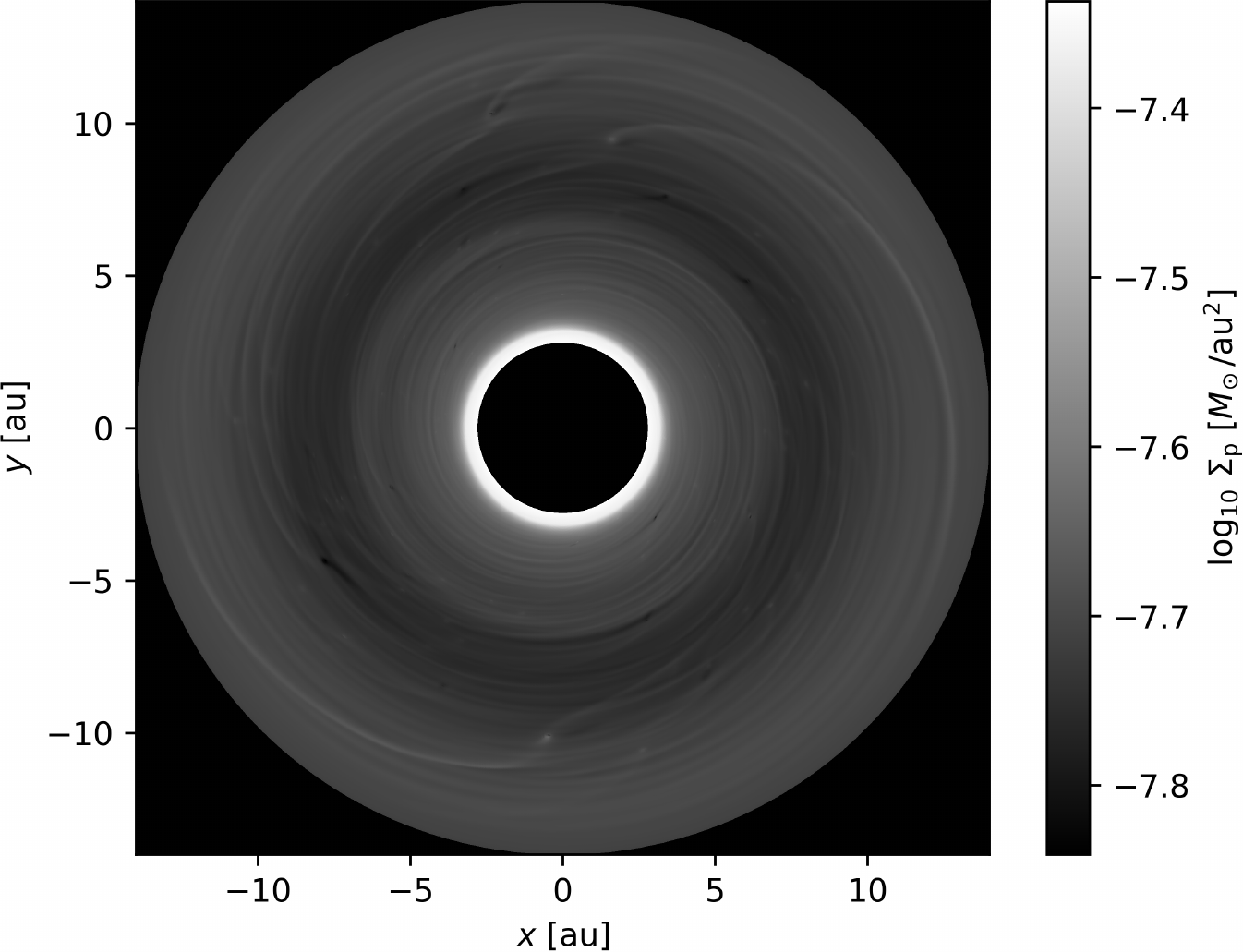}
\caption{The gas surface density~$\Sigma$ (top) and the pebble
surface density~$\Sigma_{\rm p}$ close to the end of the simulation
{\tt embryos\char`_0.1ME\char`_120}, at $t = 11200\,P_{\rm orb}$.
It is a continuation from Fig.~\ref{gasdens4_2}.
There are relatively massive embryos present in the disk,
reaching up to~$3\,M_\oplus$ (cf. the spiral arms).
They concentrate in the outer part of the disk.
In the pebble disk, one can see a darker region (low~$\Sigma_{\rm p}$) in the middle
which arises from a substantial pebble filtering by the outer embryos.
The resolution $1024\times 1536$ was used for this long-term simulation.
}
\label{gasdens448_2}
\end{figure}

The pebble filtering is closely related to a `gap' which develops in the pebble disk
(Figure~\ref{gasdens448_2}, bottom). Pebbles drifting inwards experience
a strong filtering, $\Sigma_{\rm p}$ decreases as $r\to 0$. The apparent
increase of $\Sigma_{\rm p}$ at $r\simeq 6\,{\rm au}$ is only due to
a flux conservation towards the centre.
The `winner' embryo is consequently located in the outer part of the disk,
because it experiences almost no filtering.

The separation of inclinations, or in other words a viscously stirred pebble accretion,
was studied by \cite{Levison_etal_2015Natur.524..322L}, using a Lagrangian approach.
Indeed, in our case
$\tau \simeq 0.05$,
$\alpha_{\rm p} = 10^{-4}$,
$H_{\rm p}/H = \sqrt{\alpha_{\rm p}/\tau} \simeq 0.045$,
$h = H/r \simeq 0.04$,
$h_{\rm p} = \tan I_{\rm p} \simeq 0.0018$,
i.e. a similar value as in their work.

Finally, the most massive embryos exhibit a systematic drift
towards a 0-torque radius at approximately 11\,au,
although it is somewhat hidden in Fig.~\ref{nbody.orbits.at2}.
Low-mass ones randomly walk (with a~$\!\sqrt t$ characteristic).
As soon as they reach the interaction zone of high-mass ones,
they start to randomly 'run', because the zone is very chaotic.

Luckily, the final state of the system is quite similar
to the initial conditions of the simulation \verb|CaseIII_nominal|.
There are 4 embryos with masses between $1.5$ and $3\,M_\oplus$.
They are already located close to the 0-torque radius
and they already interact with each other.
We do not think this is a mere coincidence, but rather an indication
our model can self-consistently describe both phases of evolution
in one `elegant' step.

\section{Conclusions}\label{sec:conclusions}

The simulations of the Type-I migration presented in this study
confirm that outcomes sensitively depend on the disk parameters,
as well as the initial conditions and masses of protoplanets.
We reported at least several interesting results:
  (i)~three-body encounters are needed for successful mergers;
 (ii)~in high-$\Sigma$ disks (several times MMSN) 'repulsion' events are frequent;
(iii)~a~massive coorbital pair may develop a~pebble isolation which prevents further accretion;
 (iv)~a~stabilisation and inward migration of the coorbital then occurs;
  (v)~a~dynamical tadpole torque can arise in the outer disk (as in \citealt{Pierens_2015MNRAS.454.2003P});
 (vi)~this leads to outward $\leftrightarrow$ inward migration cycles;
(vii)~the respective fast-migrating embryo may disrupt an inner coorbital pair,
solving the problem of too many coorbitals which are not observed \citep{Vokrouhlicky_Nesvorny_2014ApJ...791....6V};
(viii)~disk torques for many low-mass embryos are stochastic,
due to overlapping spiral arms, and the torques computed for single planets
(as in \citealt{Paardekooper_etal_2011MNRAS.410..293P}) are no longer valid
in this regime. This may have very important implications
for N-body models of planetary migration.

In the gas-giant zone, our simulations show a robust runaway growth
with several contributing processes, in particular
the above mentioned merging of embryos,
the Hill regime of pebble accretion which is proportional to~$M_{\rm em}$,
the pebble filtering by outer embryos, and
lower inclinations of massive embryos (being often within the pebble disk scale height),
further supporting the results of \cite{Levison_etal_2015Natur.524..322L}.

In the ice-giant zone (at 20\,au), there is a surprising convergence zone
for more massive ($5\,M_\oplus$) embryos, or mergers which may
originate from the gas-giant zone. This could possibly support
scenarios in which Neptune forms first, Uranus second, etc.
-- in an opposite way than in \cite{Isidoro_etal_2015A&A...582A..99I}.

We may be actually seeing different phases of a 'Grand Scenario',
outlined e.g. by the following sequence of simulations:
\verb|embryos_0.1ME_120|
$\rightarrow$ \verb|embryos_1.5ME_8|
$\rightarrow$ \verb|pebbleflux_2e-5|
$\rightarrow$ \verb|totmass_20ME|.
Of course, it would be better to have everything in one simulation
(and a big disk, spanning at least from the water snowline
to the outer edge, if there was any).

The time span of our longest simulations is about 150\,kyr.
If the total available mass of solids is of the order of~$130\,M_\oplus$
(as in \citealt{Levison_etal_2015Natur.524..322L}),
and the pebble flux reaches up to
$2\times 10^{-4}\,M_\oplus\,{\rm yr}^{-1}$,
its duration could be as long as 650\,kyr.
It is thus still possible to prolong our simulations using the current setup.

Ideally, one would like to continue until the gap opening in the gas disk,
which would however require
a~better treatment of the pebble isolation \citep{Bitsch_etal_2018A&A...612A..30B};
a suitable parametrisation of the gas accretion,
derived in full 3D, not in 2D \citep{Crida_Bitsch_2017Icar..285..145C};
or even until the photoevaporation,
probably including a model for the disk atmosphere \citep{Owen_etal_2011MNRAS.412...13O},
and inevitably also a planetesimal (debris) disk,
which stabilises the emerging compact planetary systems.

\section*{Acknowledgements}

The work of MB and OC has been supported by the Grant Agency of the Czech
Republic (grant no.\ 18-06083S).
The work of OC has been supported by Charles University
(research program no.\ UNCE/SCI/023;
project GA~UK no.\ 128216;
project SVV-260441).
DN's work was supported by the NASA XPR program.
Access to computing and storage facilities owned by parties and projects
contributing to the National Grid Infrastructure MetaCentrum,
provided under the programme ``Projects of Large Research, Development,
and Innovations Infrastructures'' (CESNET LM2015042), is greatly appreciated.
We also thank A. Morbidelli and E. Lega for fruitful discussions on the subject,
an anonymous referee and the editor (T.~Guillot) for constructive criticism.


\bibliographystyle{aa}
\bibliography{references}

\appendix


\section{A convergence test for $0.1M_\oplus$ embryos}\label{sec:convergence}

\begin{figure}
\centering
\includegraphics[width=9cm]{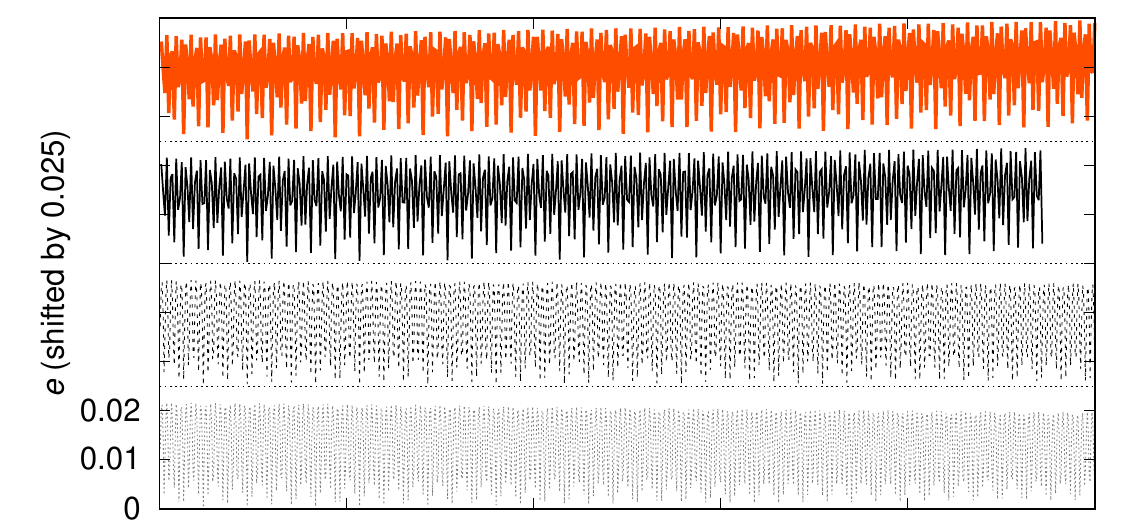}
\includegraphics[width=9cm]{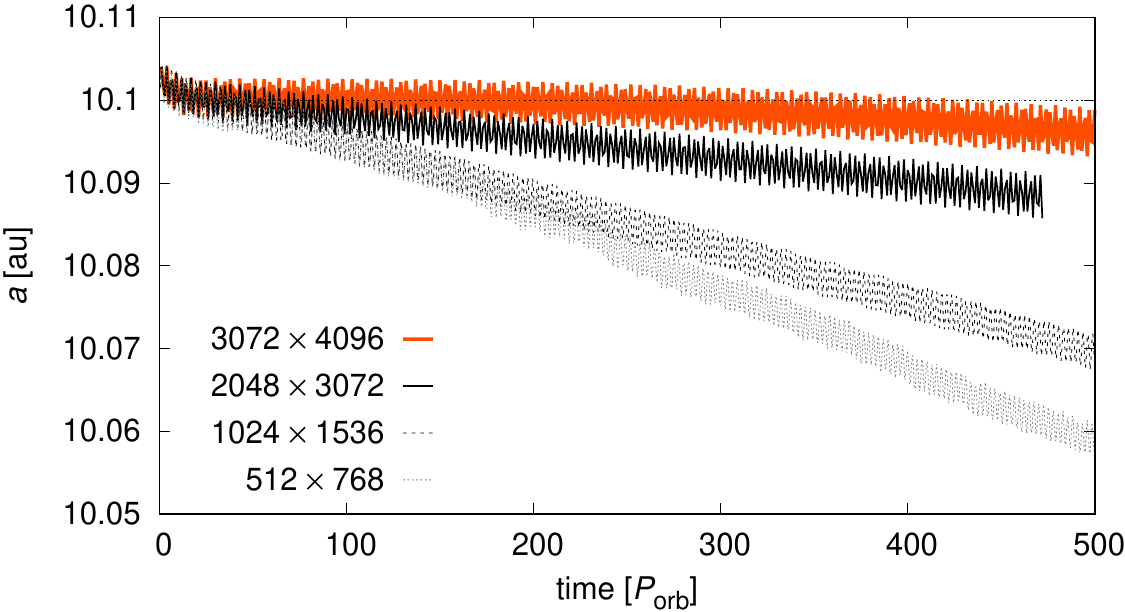}
\caption{The temporal evolution of the semimajor axis (bottom)
and the eccentricity (top) of a single low-mass $0.1\,M_\oplus$ embryo
embedded in the gas/pebble disks, for four different discretisations
in space, namely 
$512\times 768$ cells in $r$ and $\theta$ (i.e. a very low resolution),
$1024\times 1536$ (low),
$2048\times 3072$ (moderate), and
$3072\times 4096$ (high).
The amplitude eccentricity oscillations is approximately the same
for all cases, even though the torques have much larger amplitudes
for very low and low resolutions.
To obtain a correct value for the migration rate $\d a/\d t$,
one would need the moderate resolution at least. The rate $\d a/\d t$
is about three times as large for the low resolution.}
\label{test1_convergence_nbody.orbits.at}
\end{figure}

We simulated a solitary embryo with the mass $M_{\rm em} = 0.1\,M_\oplus$,
embedded in the gas/pebble disks. Practically, the setup corresponds
to the simulation \verb|embryos_0.1_120| from the main text.
We used the following resolutions (in $r,\theta$):
$512\times 768$ (very low),
$1024\times 1536$ (low),
$2048\times 3072$ (moderate),
$3072\times 4096$ (high).
Per Hill sphere, it corresponds to 1.5, 3, 5, and 8~cells in the radial direction.
The outcome is summarised in Figure~\ref{test1_convergence_nbody.orbits.at}.
The semimajor axis exbibits different migration rates,
for very low and low resolutions, they are four to three times
as large as for the moderate or high resolutions.
This is due to a poor resolution of the
Hill sphere, and the corotation region. 

On the other hand, the eccentricity oscillations have
rather similar amplitudes over the given time span
(albeit the temporal evolution is somewhat undersampled).
If we look at the torques in low-resolution simulations,
they vary with much larger amplitudes, but
on average the eccentricity seem to be sufficiently similar.
The moderate and high resolutions may exhibit a long-term trend
towards a larger asymptotic value.

Nevertheless, we decided to use $1024\times 1536$ at this stage of research,
because the computations would be otherwise prohibitively expensive
in terms of the CPU time.
We also thought the evolution may be quick enough
that the high resolution actually would not be needed
for the whole simulation time, if some mergers occur soon
for which this resolution is sufficient.

\section{Supporting figures}

\begin{figure}
\centering
\includegraphics[width=8.5cm]{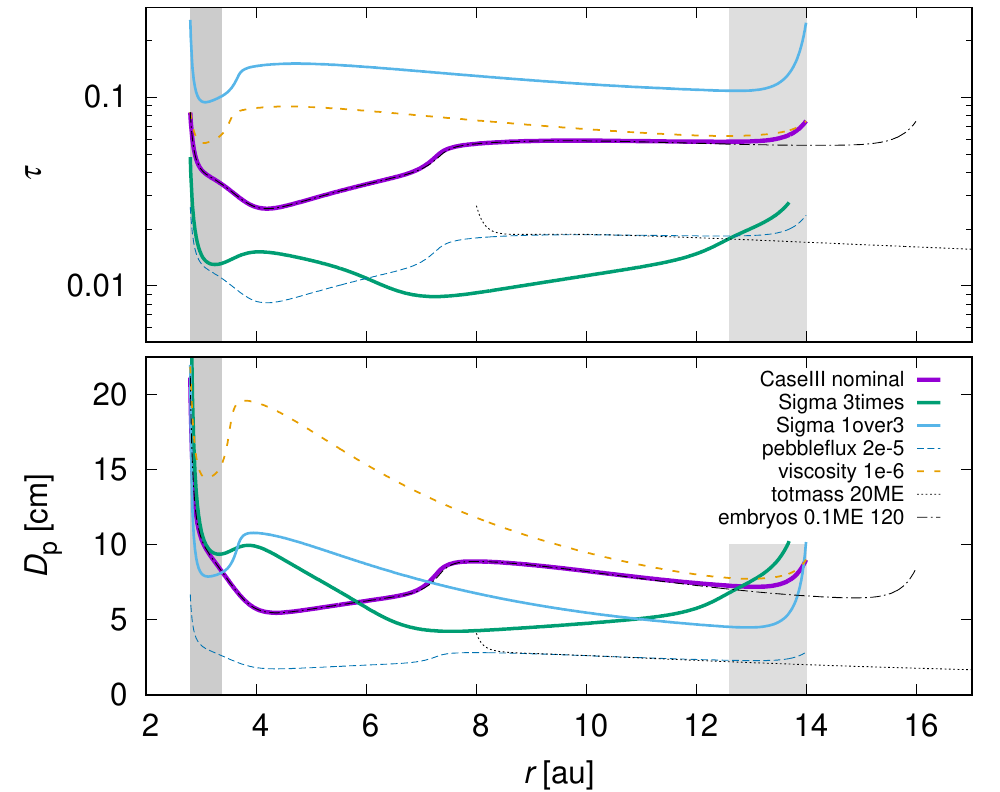}
\caption{The pebble sizes~$D_{\rm p}$ (top),
the corresponding Stokes numbers~$\tau$ (bottom)
and their dependence on the radial distance~$r$ in individual simulations.
The value of $D_{\rm p}$ changes in such a way to keep the pebble flux~$\dot M_{\rm p}$
initially constant. On the other hand, the value of $\tau$ is inversely proportional
to the coupling between the gas and pebbles.
Both quantities are increased in the damping zones (gray strips),
but this does not affect our simulation in any way.}
\label{profiles_pebble}
\end{figure}

\begin{figure}
\centering
\includegraphics[width=7cm]{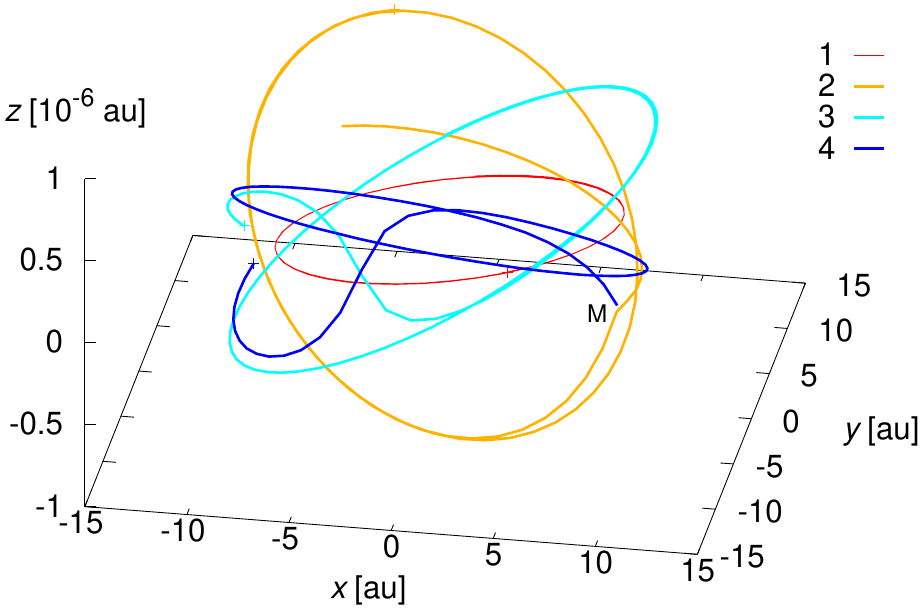}
\caption{The same as Fig.~\ref{CaseIII_nominal_Z1_MERGER1_rt} shown in 3D.
It can be seen how the encounter proceeds in the vertical direction.
Note the scale of the $z$ coordinate is very small ($10^{-6}\,{\rm au}$).
The subsequent merger event is denoted by 'M'.}
\label{CaseIII_nominal_Z1_MERGER1_xyz}
\end{figure}

\begin{figure}
\centering
\includegraphics[width=8cm]{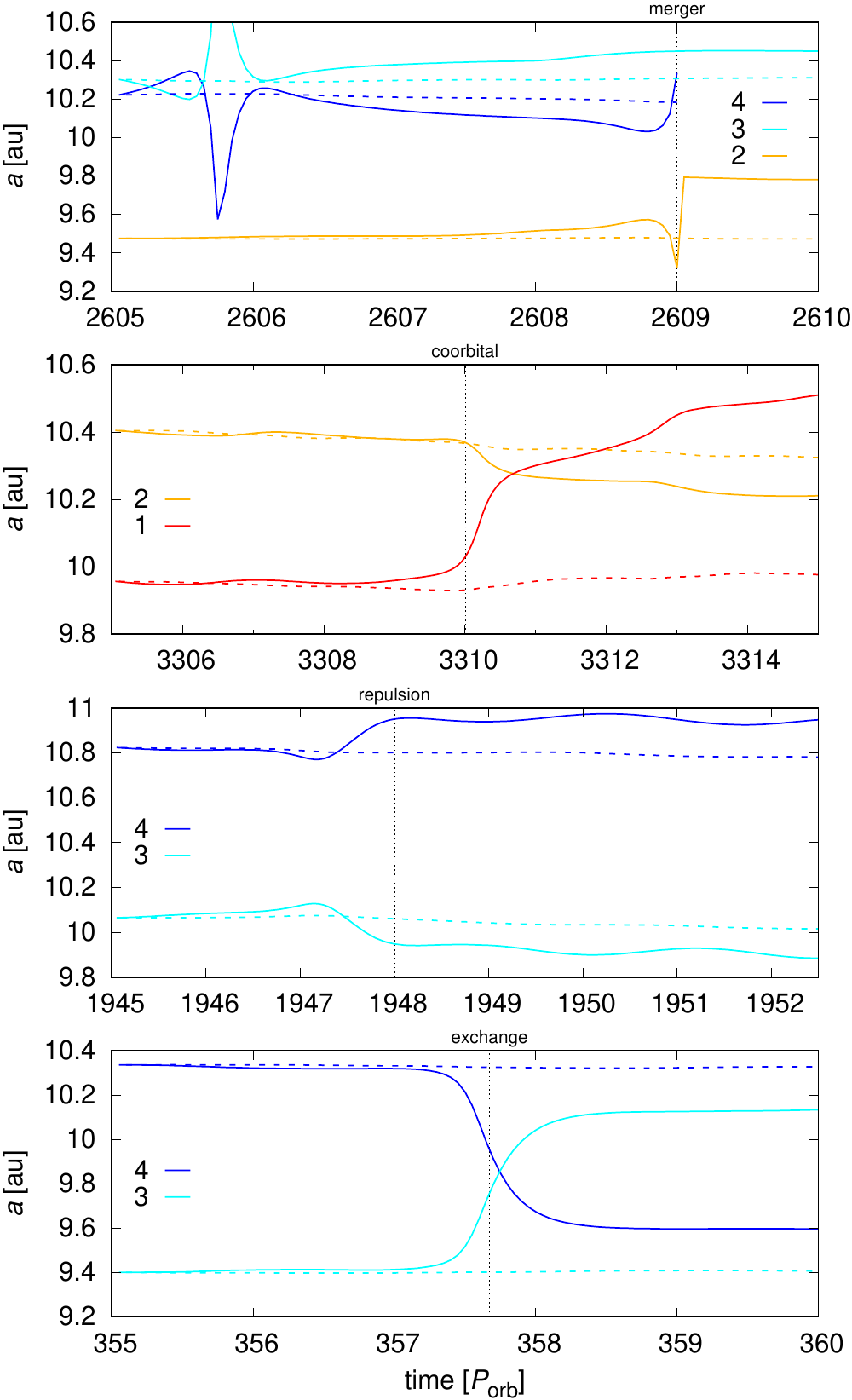}
\caption{Semimajor axis~$a(t)$ evolution corresponding to the torques~$\Gamma$ in Fig.~\ref{tqwk}.
The actual orbital evolution (solid lines) of the embryos is shown together
with a fictitious trajectory (dashed) obtained from the disk torques only,
as an integral of the first Gauss equation, $\d a/\d t = 2\Gamma/(rn)$,
where~$r$ denotes the radial distance and $n$ the mean motion.
}
\label{tqwk_integral}
\end{figure}

The supporting figures~\ref{profiles_pebble} to~\ref{tqwk_integral}
show the simulations discussed in the main text in an alternative way.

\section{Context and subsequent evolution}\label{sec:context}

In order to provide a broader context for our work and expected
time scales of a subsequent evolution we append a few remarks.
The protoplanets we focus on in this paper (with initial masses
ranging from $0.1$ to $5\,M_\oplus$) do no open a gap in the disk,
and this phase is generally called a~(fast) Type-I migration.
Its typical time scale (at the distance of Jupiter $5.2\,{\rm au}$)
is of the order of $10\,{\rm kyr}$,
after which the radial distance changes substantially, protoplanets start to interact, etc.
Only if the planetary core becomes critically massive (${\simeq}\,20\,M_\oplus$)
it accretes and repels gas from the corotation region
-- possibly in mere $10^2$ orbital periods --
and the respective torques are zeroed \citep{Lin_Papaloizou_1986ApJ...307..395L,Crida_Morbidelli_2007MNRAS.377.1324C,Crida_Bitsch_2017Icar..285..145C}.
The remaining distant spiral arms couple the planet to the disk,
which is driven by the viscosity~$\nu$,
and thus a~(slow) Type-II migration should occur,
with the time scale given by
$\left({{\rm d}a/{\rm d}t}\right)_{\rm II} = -{3\nu/(2a)}$, i.e. the order of $100\,{\rm kyr}$.

While we do not study the viscous properties of the disk in any detail,
we should note the (eddy) viscosity~$\nu$
is actually used to describe underlying (unresolved) turbulent flows,
which effectively exchange the angular momentum between differentially rotating layers.
This turbulence can be driven by the
vertical shearing instability VSI (a.k.a. Kelvin--Helmholtz in $z$ direction; \citealt{Nelson_etal_2013MNRAS.435.2610N}),
subcritical baroclinic instability SBI (essentially, Rayleigh--Taylor with heat diffusion; \citealt{Klahr_Bodenheimer_2003ApJ...582..869K}),
magneto--rotational instability MRI \citep{Balbus_Hawley_1991ApJ...376..214B,Turner_etal_2014prpl.conf..411T},
although it can be suppressed in dead zones where the ionisation is too low,
or spiral wave instability SWI (resonant coupling between spiral arms
induced by an embedded planet and inertial-gravity waves; \citealt{Bae_etal_2016ApJ...833..126B}).
Theoretically, it is also possible that the accretion inflow is driven by
a~stellar wind \citep{Gunther_2013AN....334...67G,Turner_etal_2014prpl.conf..411T},
if it was not weak,
X-wind at the disk edge \citep{Shu_etal_1994ApJ...429..781S},
or by magneto-centrifugal wind and loading of ions \citep{Anderson_etal_2005ApJ...630..945A,Suzuki_etal_2016A&A...596A..74S},
which all can affect the disk {\em surface\/} and carry some part of the angular momentum away.

From the observational point of view, a full disk should become pre-transitional
(as defined by \citealt{Espaillat_etal_2014prpl.conf..497E}) in the course of the outlined evolution.
Transitional or even evolved disks correspond to much later phases.
In the framework of young stellar object (YSO) classification,
based on $2.2\hbox{ to }20\,\mu{\rm m}$ spectral slope of $\lambda F_\lambda$,
it corresponds to Class~II objects with moderately negative slope.
The above-mentioned gap opening should eliminate a part of
the disk with moderate temperatures and result in a pronounced
decrease of spectral-energy distribution around $\lambda \simeq 10\,\mu{\rm m}$.

At the same time, the central star (Sun) irradiating the disk is a pre-main
sequence object of T~Tauri type, for which we assume parameters
$M_\star = 1\,M_\odot$,
$T_{\rm eff} \doteq 4\,300\,{\rm K}$,
$R_\star \doteq 1{,}5\,R_\odot$ (\citealt{Paxton_etal_2015ApJS..220...15P}).
It is usually expected the star had already low accretion rate
(of the order of $10^{-8}\,M_\odot\,{\rm yr}^{-1}$; \citealt{Ingleby_etal_2013ApJ...767..112I}).
In the course of stellar evolution, the convective zone
recedes away from the core and only a subsurface convection remains.
Most likely, a shearing zone (tachocline) is then formed
which is related to an onset of the solar dynamo,
an increase of the FUV, X-ray flux, and
consequent photoevaporation of the disk \citep{Owen_etal_2011MNRAS.412...13O}.
It would take another ${\simeq}\,26\,{\rm Myr}$ to reach the zero-age main sequence (ZAMS).
This is a typical situation for all F, G, K stars;
more massive stars would be classified as Herbig Ae, Be objects.

\label{lastpage}

\end{document}